\documentclass[11pt,tightenlines,eqsecnum,floats,aps,amsmath,amssymb,nofootinbib,prd,shownopacs,floatfix]{revtex4}

\usepackage{amsmath,amssymb,amsfonts,dcolumn,color,graphicx,graphics,latexsym}
\usepackage[mathscr]{eucal}
\usepackage[latin1]{inputenc}
\usepackage{enumerate}
\newcommand{\be}{\begin{equation}}
\newcommand{\ee}{\end{equation}}
\newcommand{\ba}{\begin{eqnarray}}
\newcommand{\ea}{\end{eqnarray}}

\usepackage{epstopdf}\usepackage{epstopdf}
\usepackage{latexsym}
\usepackage{amssymb}
\usepackage{amsmath}
\usepackage{color}
\usepackage{float}
\usepackage{mathrsfs}
\usepackage{multirow}
\usepackage{tabularx,booktabs}
\usepackage{latexsym}
\usepackage{amssymb}
\usepackage{amsmath}
\usepackage{color}
\usepackage{float}
\usepackage{mathrsfs}
\usepackage{multirow}
\usepackage{tabularx,booktabs}

\newcommand{\bmult}{\nopagebreak[3]\begin{multline}}
\newcommand{\emult}{\end{multline}}
\newcommand{\nb}{\nonumber}
\def\d{{\rm d}}

\newcommand{\rc}{\rho^{\scriptscriptstyle{\mathrm{I}}}_c}

\begin{document}

\title{Generic absence of strong singularities and geodesic completeness in modified loop quantum cosmologies}
\author{Sahil Saini}
\email{ssaini3@lsu.edu}
\author{Parampreet Singh}
\email{psingh@lsu.edu}
\affiliation{ Department of Physics and Astronomy,\\
	Louisiana State University, Baton Rouge, LA 70803, U.S.A.}

\begin{abstract}
Different regularizations of the Hamiltonian constraint in loop quantum cosmology yield modified loop quantum cosmologies, namely mLQC-I and mLQC-II, which lead to qualitatively different Planck scale physics. We perform a comprehensive analysis of resolution of various singularities in these modified loop cosmologies using effective spacetime description and compare with earlier results in standard loop quantum cosmology. We show that the volume remains non-zero and finite in finite time evolution for all considered loop cosmological models. Interestingly, even though expansion scalar and energy density are bounded
due to quantum geometry, curvature invariants can still potentially diverge due to pressure
singularities at a finite volume. These divergences are shown to be harmless since geodesic
evolution does not break down and no strong singularities are present in the effective spacetimes of loop cosmologies. Using a phenomenological matter model, various types of exotic strong and weak singularities, including big rip, sudden, big freeze and type-IV singularities, are studied. We show that as in standard loop quantum cosmology, big rip and big freeze singularities are resolved in mLQC-I and mLQC-II, but quantum geometric effects do not resolve sudden and type-IV singularities. 
\end{abstract}

\maketitle

\section{Introduction}

Singularities are generic features of Einstein's theory of general relativity (GR) and are associated with a break down of geodesics and potential divergences in physical quantities such as the energy density, the expansion scalar and the curvature invariants. However, not all events characterized by these divergences necessarily reflect the breakdown of geodesic evolution or impossibility of survival of sufficiently strong detectors. The criterion of strong versus weak singularities is generally used to distinguish the problematic divergences representing boundaries of the spacetime from harmless events \cite{ellis1977,tipler1977singularities,krolak,ck-1985conditions}. A strong singularity is characterized by the property that it completely destroys all in-falling objects regardless of their properties. Singularities that do not satisfy this criterion are called weak singularities, and a strong enough object could go past them without being completely destructed. It is conjectured that all curvature singularities which result in geodesic incompleteness are strong singularities \cite{krolak}.  Well known examples of singularities in GR, namely the central singularity in Schwarzschild black holes and the big bang/big crunch are strong singularities. In recent years, various other examples of singularities in cosmological models have been found. These include big rip \cite{bigrip}, sudden singularities \cite{sudden}, big freeze \cite{Nojiri2005,freeze} and type-IV (or generalized sudden) singularities \cite{Nojiri2005,type4}. Of these, big rip and big freeze are known to be strong singularities in GR, where as sudden and type-IV singularities are weak singularities \cite{Jambrina1}.

It has been long expected that quantum gravitational modifications to GR will provide insights on resolution of classical singularities. In the last decade, this expectation has turned out to be true in loop quantum cosmology (LQC) which is a non-perturbative quantization of  cosmological spacetimes based on techniques of loop quantum gravity (LQG) \cite{as-status}. One of the main results from LQC is the resolution of big bang/big crunch singularities at the quantum level which finds its roots in the underlying quantum geometry of LQG. The big bang is replaced by a quantum bounce when the energy density reaches Planckian regime, and the expanding branch of our universe is connected to a contracting pre-bounce branch \cite{aps1,aps3,apsv}. At the quantum level, the expectation values of the energy density operator have a universal maximum in the physical Hilbert space \cite{slqc}, and the quantum probability for the singularity  to occur turns out to be vanishing \cite{craig-singh}. The evolution in LQC is dictated by the quantum Hamiltonian constraint, a difference equation, which interestingly can be captured to an excellent approximation by an effective spacetime description for isotropic as well as anisotropic spacetimes \cite{numlsu}. Using this effective description various phenomenological implications have been studied \cite{agullo-singh}, and generic resolution of cosmological singularities in isotropic and anisotropic models has been understood in detail \cite{ps09,psfv,ps11,ps14,ks-strong,b2-strong,b9-strong}, with ongoing attempts to include inhomogeneities \cite{madrid}. These studies establish that strong singularities are generically resolved in LQC, whereas weak singularities are generally ignored by quantum geometry.\footnote{Exceptions however have been found for spatially curved isotropic model, and Bianchi-I flat Kasner spacetime where quantum geometry effects resolve even weak singularities \cite{psfv,kasner-flat}.}

Despite above progress in understanding resolution of singularities in LQC, one of the open questions is how robust are these results to quantization ambiguities such as of choosing different regularizations of the Hamiltonian constraint motivated from LQG. 
Here let us note that in the standard LQC, the Euclidean and the Lorentzian terms in the Hamiltonian constraint are combined together before quantization because owing to homogeneity both the terms are proportional to each other. The subsequent quantization led to the replacement of the big bang singularity by a symmetric bounce connecting the current expanding classical universe to a pre-bounce contracting classical universe \cite{aps1,aps3,slqc,cs-recall}. If the Lorentzian term is treated independently for loop quantization using Thiemann's regularization \cite{Thiemann}, one is led to two modified versions of standard LQC -- mLQC-I and mLQC-II (using the conventions of \cite{Li2018b} where mLQC stands for modified-LQC in contrast to standard LQC \cite{as-status} which will be often denoted as LQC in the manuscript). The two variants of LQC, mLQC-I and mLQC-II, differ from each other in the way Lorentzian term is quantized. While the extrinsic curvature in the Lorentzian term is directly quantized in mLQC-I using Thiemann's identities involving holonomies of the connection, mLQC-II exploits the symmetries of the spatially flat FLRW model to first replace the extrinsic curvature in the Lorentzian term by the Ashtekar-Barbero connection before quantization \cite{Yang2009}. 

Both of these variants of standard LQC -- mLQC-I and mLQC-II,  lead to quantum Hamiltonian constraints as fourth order quantum difference equations, in contrast to standard LQC where the difference equation is of second order \cite{von-Neumann}. Higher order difference equation can result also in LQC by considering higher representation for SU(2) than the fundamental representation in the quantization of Hamiltonian constraint even without treating Lorentzian term independent of the Euclidean term \cite{brahma}\footnote{Note that early works in LQC sometimes considered higher spin representations only to regularize inverse volume operator in the matter part, which does not lead to a higher order difference equation (see for eg. \cite{higherj}).} (see also Ref. \cite{perez} for an earlier work in LQG). The initial value problem for these quantum difference equations is more involved than in standard LQC. In particular, for mLQC-I and mLQC-II one needs to provide the value of the wavefunction and its first three derivatives at an initial time with respect to a suitable relational clock, such as the massless scalar field $\phi$ used in LQC. One can then use the procedure used often for higher order initial value problems by converting the fourth order initial value problem to four first order initial value problems and solving them using standard techniques (see for eg. \cite{bradie}). Our goal in this manuscript is focused on the effective dynamics obtained from the effective Hamiltonian which has non-trivial changes in comparison to standard LQC, resulting in a complicated phase space structure which consists of two distinct branches of solutions. These two branches couple in a non-trivial way in mLQC-I, and of the two only one is physically relevant for mLQC-II \cite{Li2018,Li2018b}. We note that treating Lorentzian term independent of Euclidean term was first explored  Ref. \cite{Bojowald2002}, where mLQC-I in the $ \mu_o$ like scheme of \cite{aps1} was obtained. Later in Ref. \cite{Yang2009}, both mLQC-I and mLQC-II were obtained for the improved dynamics ($ \bar{\mu} $ scheme) employed in Ref. \cite{aps3}. Recently, the effective Hamiltonian for mLQC-I has also been obtained from a cosmological sector of LQG using complexifier coherent states for the $\mu_o$ scheme \cite{Dapor2017}.

Above different ways to regularize Hamiltonian constraint in LQC result in qualitative differences in Planck scale physics. For instance, the bounce turns out to be asymmetric in mLQC-I, with the asymptotic pre-bounce branch resembling a de-Sitter spacetime with an effective cosmological constant of Planckian size and a different value of Newton's gravitational constant \cite{Li2018}. Phenomenological investigations of mLQC-I effective dynamics using the modified Friedmann and Raychaudhuri equations  show that both the phase space branches of mLQC-I are necessary to describe the phase space evolution due to the effective Hamiltonian. The two branches, having separate Friedmann and Raychaudhuri equations, are connected at the bounce for mLQC-I leading to an asymmetric bounce in mLQC-I. Numerical investigations in mLQC-I using sharply peaked Gaussian states have confirmed these predictions of the effective dynamics showing a high degree of agreement between the effective spacetime description and the quantum dynamics in case of mLQC-I \cite{adlp2018}. In contrast to mLQC-I where a non-trivial union is necessary between two branches, one of the branches in case of mLQC-II turns out to be unphysical, and only one branch (and the corresponding one set of Friedmann and Raychaudhuri equations) describes the full phase space evolution generated by the effective Hamiltonian \cite{Li2018b}. The bounce in mLQC-II is symmetric as in standard LQC, but unlike in LQC where the modified Friedmann and Raychaudhuri equations contain  $\rho^2$ terms, one obtains higher order corrections. The dependence on higher order terms is similar as in mLQC-I. Notably the bounce densities in LQC, mLQC-I and mLQC-II all turn out to be different. 

Various aspects of mLQC-I and mLQC-II effective dynamics and their detailed phenomenological implications are being studied. Detailed investigations of the properties of effective dynamics and inflationary attractors for various potentials have been carried out \cite{Li2018,Li2018b}, power spectrum of the cosmic microwave background using scalar perturbations \cite{power-spectrum} has been studied, and the bounce in mLQC-I has also been studied from a dynamical systems point of view \cite{haro}.  At the level of difference equations, von-Neumann stability analysis of mLQC-I and mLQC-II reveals various interesting features \cite{von-Neumann} in synergy with observations in Ref. \cite{Li2018,Li2018b}. Our goal in this manuscript is to perform an analysis of generic resolution of singularities in mLQC-I and mLQC-II for arbitrary minimally coupled matter. The resolution of big bang singularities in these models for massless scalar field and inflationary potentials has been earlier studied in \cite{Li2018,Li2018b}, but results so far tell little about the way singularities are resolved in general for arbitrary matter and the way situation compares with previous results in LQC.

We show in our analysis that the volume remains non-zero and finite for all finite time evolution and the Hubble rate or the expansion scalar, and the energy density are generically bounded in both mLQC-I and mLQC-II, as is the case for LQC. However, the time derivative of the Hubble rate and as a result curvature invariants can diverge if the pressure diverges at a finite value of energy density. The phase space is maximally extendible except for such pressure divergences which are shown to be harmless as the geodesics remain regular at such events. Further, using the necessary and sufficient conditions for the existence of strong singularities, we show that while strong singularities are present in the classical dynamics, they are absent from the effective spacetime of mLQC-I and mLQC-II models. Potential pressure singularities thus amount to weak singularities. 

After obtaining above results for a general minimally coupled matter, we numerically investigate the fate of exotic singularities using phenomenological matter considered in Ref. \cite{Nojiri2005} which allows various types of singularities. Using convention of Ref. \cite{Nojiri2005}, these singularities are labeled as type-I, type-II, type-III and type-IV singularities. In type-I singularities, the scale factor, the energy density and the pressure diverge in a finite time evolution. These are the big rip singularities. Type II, i.e. sudden singularities, occur when the pressure diverges at a finite value of energy density, which causes divergence in the Ricci scalar and the break down of phase space evolution. In type-III (or big freeze) singularities,  the energy density, the pressure and the Hubble rate diverge at a finite value of the scale factor. In type-IV, i.e. the generalized sudden singularities, it is the time derivatives of the curvature invariants such as Ricci scalar that diverge while the energy density, the pressure and the Ricci scalar remain finite, and the phase space evolution does not break down. Of these, type-I and type-III are strong curvature singularities while type-II and type-IV are of weak curvature type \cite{ps09}. Using effective dynamics of mLQC-I and mLQC-II, we analyze the fate of these singularities. 
 We find that type-I and type-III singularities  are resolved in mLQC-I and mLQC-II, as in LQC, although the details differ from one to another. The type-II and type-IV singularities are not resolved in mLQC-I and mLQC-II. This is not problematic as these singularities are weak singularities and the geodesics are well behaved and extendible at these events. We also find that the dynamics of mLQC-II is qualitatively similar to that of standard LQC while mLQC-I differs from both of these qualitatively in the pre-bounce branch.

The manuscript is organized as follows. In Sec. II, we preview the basic Hamiltonian dynamics of classical FLRW model, and the effective Hamiltonian dynamics of the standard LQC, mLQC-I and mLQC-II. In Sec. III, we analyze the behavior of geodesics at the classically singular events and analyze the strength of singularities. We show that in all loop cosmologies considered in this manuscript the volume remains non-zero in finite time evolution and energy densities are bounded above. But, curvature invariants can still diverge. In Sec. IV we consider a particular matter equation of state to numerically analyze the type-I, type-II, type-III and type-IV singularities in mLQC-I and mLQC-II and compare with classical theory and LQC.  We conclude  with a summary of results obtained in Sec. V.

\section{Effective Hamiltonian dynamics}
In this section, we review the basic elements of the effective dynamics of three different loop cosmological models: LQC and its two modifications mLQC-I and mLQC-II resulting from different treatment of the Lorentzian term in the spatially flat FLRW spacetime. We compare and contrast the effective dynamics of these models among themselves and also with the classical GR description. This is intended to provide a background for proving analytical results in the next section regarding geodesic completeness and absence of strong singularities in the effective loop cosmological spacetimes. This section also provides dynamical equations for the numerical investigations in Sec. IV to phenomenologically study strong and weak singularities. 

In the following, the first subsection reviews the classical Hamiltonian dynamics of the FLRW spacetime. Expressions for some important quantities such as the energy density, Hubble rate, expansion scalar, Ricci scalar and the time derivatives of the Hubble rate and Ricci scalar are obtained in classical cosmology using Hamiltonian dynamics. In Sec. IIB, we revisit the effective dynamics of the standard LQC for spatially flat isotropic cosmology and obtain the corresponding expressions for all the relevant quantities for comparison with classical dynamics. We recall results of Ref. \cite{ps09} showing boundedness of various quantities such as the energy density and the Hubble rate, and possible divergences in Ricci scalar and other curvature invariants. In the next two subsections, we repeat the analysis for two variations of LQC: mLQC-I \cite{Yang2009,Dapor2017,Li2018} and mLQC-II \cite{Yang2009,Li2018b}, and discuss bounds on energy density, Hubble rates and potential curvature pathologies. For all loop cosmological models, we show that the scale factor or the volume does not vanish in finite time evolution. This result along with boundedness of Hubble rate turns out to be crucial to prove geodesic completeness in Sec. III.

\subsection{Classical Dynamics}
We consider the spacetime manifold for the spatially flat FLRW model as $ \mathbb{R} \times \Sigma $ where $ \Sigma=\mathbb{R}^3 $. The spacetime can be locally described by the metric,
\begin{equation}\label{key}
\mathrm{d}s^2= - N(t)^2 \mathrm{d}t^2 + a(t)^2 (\mathrm{d}r^2 + r^2 (\mathrm{d}\theta^2 + \sin^2 \theta \mathrm{d}\phi^2)),
\end{equation} 
where $ a(t) $ is the scale factor of the universe. We will choose the lapse $ N=1 $ in the following analysis. The isotropy and homogeneity of the spacetime ensures we only have one degree of freedom which is contained in the scale factor. The dynamics of this spacetime can be reformulated in a canonical form using the Hamiltonian formulation of GR. The Hamiltonian so obtained for spatially flat FLRW spacetime is given by,
\begin{equation}\label{key}
\mathcal{H_\mathrm{cl}} = -\frac{3 b^2 v}{8\pi G \gamma^2} + \mathcal{H_\mathrm{m}}
\end{equation}
where we have used variables $b$ and $v$ commonly used in LQC literature, and $\gamma$ is the Barbero-Immirzi parameter. Here, the variable $ v=a^3 $ represents the physical volume of a fiducial cell whose coordinate volume is chosen to be unity. And, the conjugate variable $ b$ is proportional to the Hubble rate $H = \dot a/a$: $b=\gamma \dot a/a $ on the physical solutions of GR. %Here $ \gamma $ is the Barbero-Immirzi parameter which using black hole thermodynamics will be set to $\gamma \approx 0.2375$. 
The matter Hamiltonian $ \mathcal{H_\mathrm{m}} $ is assumed to represent non-dissipative minimally coupled matter.  The phase space variables satisfy 
\begin{equation}
\lbrace b,v \rbrace = 4\pi G \gamma  \label{pbracket},
\end{equation}
which leads to the following equations of motion,
\begin{equation} \label{bcl}
\dot b = 4 \pi G \gamma \frac{\partial \mathcal{H_\mathrm{cl}} }{\partial v} = - \frac{3b^2}{2\gamma} - 4\pi G\gamma P, ~~~~\mathrm{and} ~~~
\dot v = - 4 \pi G \gamma \frac{\partial \mathcal{H_\mathrm{cl}} }{\partial b} = \frac{3bv}{\gamma}.
\end{equation}
Here $ P $ is the isotropic pressure defined by $ P:=-\frac{\partial \mathcal{H_\mathrm{m}}}{\partial v} $. 

The vanishing of the Hamiltonian constraint, $\mathcal{H_\mathrm{cl}} \approx 0$ implies that the energy density, $\rho = {\cal H}_m/v$, is dynamically equal to,
\begin{equation}\label{key}
\rho = \frac{3 b^2}{8\pi G \gamma^2}.
\end{equation}
In the classical theory, divergence in energy density can thus be seen directly in terms of the divergence in $b$ or the Hubble rate. The above equation when expressed in terms of the Hubble rate yields the classical Friedmann equation:
\begin{equation}\label{key}
H^2 = \frac{8\pi G}{3} \rho.
\end{equation}
The Raychaudhuri equation can be obtained using $\dot b$, which turns out to be   
\begin{equation}
\frac{\ddot a}{a} = H^2 + \dot H = - \frac{4\pi G}{3} (\rho + 3P), \label{Raychaudhari_FRW_classical}
\end{equation}
using which we obtain the following relation between the Ricci scalar, energy density and pressure in the classical theory:
\begin{equation}
R = 6 \bigg(H^2 + \frac{\ddot a}{a}\bigg) = 8\pi G (\rho - 3P).
\end{equation}
It can be seen that the Friedmann and Raychaudhuri equations lead to the energy conservation equation $ \dot\rho = -3H(\rho +P) $. It is also useful to consider the time derivative of the Ricci scalar $ \dot R $ which turns out to be useful to understand type-IV singularities in Sec. IV.  The time derivative of the Ricci scalar can be shown to be $ \dot R = 6(\ddot H + 4H \dot H) $ where,
\begin{equation}
\dot H = -4\pi G (\rho + P).
\end{equation}

With above equations, we need the equation of state $ P=P(\rho) $ to evaluate the time derivative of the pressure $ P $ in order to evaluate $ \dot R $. We will see concrete examples of this in Sec. IV where we consider a particular equation of state and perform numerical simulations for different types of singularities. For matter, satisfying weak energy condition above equations result in a big bang singularity in the past evolution where energy density, pressure, Ricci scalar and its time derivative diverge in a finite time. At this instant of time, the scale factor or the volume of the universe vanishes causing geodesic incompleteness. 

\subsection{Effective dynamics in standard LQC}

Our purpose in this subsection is to recall the effective dynamics of standard LQC description of the spatially flat FLRW spacetime in the $ \bar{\mu} $ quantization obtained in \cite{aps3}. For details we refer the reader to Ref. \cite{ps09} (see also Ref. \cite{cs09}). In this quantization,  the Lorentzian and Euclidean terms are combined into a single term at the classical level and then quantized. The effective Hamiltonian for the FLRW spacetime in standard LQC is given by:
\begin{equation}
\mathcal{H}=-\frac{3v}{8\pi G \gamma^2 \lambda^2} \sin^2(\lambda b) + \mathcal{H_\mathrm{m}}, \label{Hamiltonian_stdLQC}
\end{equation}
where $ \lambda = \sqrt{\Delta} $. The constant $ \Delta = 4 \sqrt{3} \pi \gamma \ell_{\mathrm{Pl}}^2 $ is the minimum area eigenvalue in LQG. The Hamilton's equations for $b$ and $v$ are,
\begin{eqnarray} \label{blqc}
\dot b &=& - \frac{3 \sin^2(\lambda b)}{2 \gamma \lambda^2} -4\pi G\gamma P, \\
\dot v &=& \frac{3 \sin(2\lambda b)}{2 \gamma \lambda} v .
\end{eqnarray}
Provided that the pressure $ P $ does not diverge, the above equations are a coupled system of linear first order ordinary differential equations (ODEs) with bounded coefficients, implying that the phase space is maximally extendible except for pressure divergences \cite{vrabie}. Specifically, the equation for volume can be integrated to yield,
\begin{equation}\label{key}
v(t) = v_o \exp \bigg[ \int_{t_o}^t \frac{3 \sin(2\lambda b)}{2 \gamma \lambda} \mathrm{dt} \bigg],
\end{equation}
where $ v_o $ is the value of the volume at some initial time $ t_o $. Since the integrand is a bounded function, $ v $ and consequently the scale factor $ a $ remain finite and non-zero for all finite time evolution. 

From the vanishing of the effective Hamiltonian constraint \eqref{Hamiltonian_stdLQC}, the energy density is dynamically equal to,
\begin{equation}\label{rholqc}
\rho = \frac{3}{8\pi G \gamma^2 \lambda^2} \sin^2(\lambda b).
\end{equation}
We note that energy density in standard LQC is generically bounded in contrast to classical dynamics. This can be used to express the Hubble rate, the modified Raychaudhuri equation and the Ricci scalar in terms of the energy density as follows,
\begin{eqnarray}
H^2 &=& \frac{8\pi G}{3} \rho \bigg(1-\frac{\rho}{\rho_{c}}\bigg), \\
\frac{\ddot a}{a} &=& - \frac{4\pi G}{3}\rho \bigg(1-4\frac{\rho}{\rho_{c}}\bigg) -4\pi G P \bigg(1-2\frac{\rho}{\rho_{c}}\bigg), \\
R &=& 8\pi G \bigg(\rho-3P+2\frac{\rho}{\rho_{c}}(\rho+3P)\bigg),
\end{eqnarray}
where $ \rho_{c} = 3/(8\pi G \gamma^2 \lambda^2) \approx 0.41 \rho_{\mathrm{Pl}} $. We note from the above equations that the Hubble rate and hence the expansion scalar $\theta = 3 H$ is generically bounded. Even though energy density is bounded above as seen from eq. (\ref{rholqc}), the second derivative of the scale factor and the Ricci scalar can still diverge if pressure $ P $ diverges at a finite value of $\rho$. We can use $ \dot R = 6(\ddot H + 4H \dot H) $ to find the time derivative of the Ricci scalar where $ \dot H $ is given by,
\begin{eqnarray}\label{key}
\dot H = -4\pi G (\rho + P)\bigg(1-2\frac{\rho}{\rho_c}\bigg).
\end{eqnarray}
It is straightforward to see that  $ \dot{R} $ can diverge even if the energy density, pressure, the Hubble rate and the Ricci scalar are bounded but the time derivative of the pressure $ \dot{P} $ diverges at finite value of the pressure. We will show in the next section that  divergences such as these and those where Ricci scalar diverges at a finite value of energy density can only lead to weak curvature singularities that are harmless and geodesics can be extended beyond them. 

\subsection{Effective dynamics of mLQC-I}
Unlike in standard LQC, where the Euclidean and Lorentzian terms are treated at equal footing and combined before quantization, one can treat Lorentzian term independently which results in a modification of the Hamiltonian constraint resulting in changes to   Planck scale physics. The first of these choices, first studied in Ref. \cite{Yang2009} and recently re-examined in Ref. \cite{Dapor2017} arises from expressing the extrinsic curvature in the Lorentzian term in terms of holonomies using Thiemann's regularization \cite{Thiemann}. The resulting quantization leads to a significantly different pre-bounce physics which turns out to be a de-Sitter phase but with a renormalized Newton's constant \cite{Li2018}.\footnote{A similar example of a quantum emergent spacetime with similar peculiar properties has been earlier found in LQC for Kantowski-Sachs spacetimes \cite{djs}.} The bounce is asymmetric even for the case of the massless scalar field which yields a symmetric bounce in LQC.   The modified Friedmann and Raychaudhuri equations are far more involved than in standard LQC, containing higher order terms than the quadratic modification in standard LQC \cite{Li2018}. Further, unlike standard LQC there exist two branches of solutions in the effective phase space which need to be finely matched at the bounce for a consistent picture of dynamics \cite{Li2018}.

The effective Hamiltonian in mLQC-I is given by \cite{Yang2009,Li2018}:
\begin{equation}
\mathcal{H}=\frac{3v}{8\pi G \lambda^2} \bigg[\sin^2(\lambda b) - \frac{(\gamma^2 +1) \sin^2(2\lambda b)}{4\gamma^2} \bigg] + \mathcal{H_\mathrm{m}},
\end{equation}
which yields the following  Hamilton's equations
\begin{eqnarray}\label{bmlqc-I}
\dot b &=& \frac{3 \sin^2(\lambda b)}{2 \gamma \lambda^2} \bigg[\gamma^2 \sin^2(\lambda b) - \cos^2(\lambda b)\bigg] - 4\pi G \gamma P, \label{EoM_asy_modLQC1} \\
\dot v &=& \frac{3 v \sin(2\lambda b)}{2 \gamma \lambda} \bigg[(\gamma^2 +1) \cos(2\lambda b) - \gamma^2\bigg]. \label{EoM_asy_modLQC2}
\end{eqnarray}
As in standard LQC, here also the phase space is maximally extendible for finite time evolution except for pressure divergences as the Hamiltonian's equations are linear first order ordinary differential equations with bounded coefficients except for the pressure $ P $. Upon integration of the equation for volume, we get
\begin{equation}\label{key}
v(t)= v_o \exp \left\lbrace \int^t_{t_o} \mathrm{dt} \frac{3 \sin(2\lambda b)}{2 \gamma \lambda} \bigg[(\gamma^2 +1) \cos(2\lambda b) - \gamma^2\bigg] \right\rbrace .
\end{equation}
Since the integrand of the exponential is an always bounded function, the volume (and hence also the scale factor) remains finite and non-zero for all finite time evolution.

The energy density is given by the vanishing of the effective Hamiltonian constraint, which is 
\begin{equation}\label{key}
\rho = -\frac{3}{8\pi G \lambda^2} \bigg[\sin^2(\lambda b) - \frac{(\gamma^2 +1) \sin^2(2\lambda b)}{4\gamma^2} \bigg].
\end{equation}
We note that the energy density is generically bounded for all times as in standard LQC. However, in contrast to standard LQC, the expression for energy density here is a quadratic equation for $ \sin^2 (\lambda b) $ which then leads to two allowed values of $ b $ in terms of the energy density. They have been named the $ b_{-} $ and $ b_{+} $ branches and investigated in detail  in \cite{Li2018}. As shown in Ref. \cite{Li2018}, it is interesting to see the way existence of these two branches complicates the phase space evolution and leads to an asymmetric bounce. These two roots lead to different expressions for the modified Friedmann and the Raychaudhuri equations, and it turns out that both of them are needed to span the whole evolution provided by equations of motion \eqref{EoM_asy_modLQC1} and \eqref{EoM_asy_modLQC2}. Each root spans a part of the solution and they are connected at the point where the energy density reaches the bounce density.\footnote{Such branches for higher order modifications of LQC were earlier studied in Ref. \cite{ss-ps}.} Following the convention of Ref. \cite{Li2018}, we denote the bounce density in mLQC-I as $\rho^{\scriptscriptstyle{\mathrm{I}}}_c$ to distinguish from bounce density $\rho_c$ in LQC. We briefly discuss relevant equations for these two branches in the following. 

\subsubsection{The $ b_{-} branch $}

In this case the modified Friedmann and Raychaudhuri equations are \cite{Li2018},
\begin{eqnarray}
H^2 &=&\frac{8\pi G \rho}{3}\left(1-\frac{\rho}{\rho^{\scriptscriptstyle{\mathrm{I}}}_c}\right)\Bigg[1  +\frac{\gamma^2}{\gamma^2+1}\left(\frac{\sqrt{\rho/\rho^{\scriptscriptstyle{\mathrm{I}}}_c}}{1 +\sqrt{1-\rho/\rho^{\scriptscriptstyle{\mathrm{I}}}_c}}\right)^2\Bigg],\\
\frac{\ddot a}{a} &=&-\frac{4\pi G}{3}\left(\rho + 3P\right)
  + \frac{4\pi G \rho}{3}\left[\frac{\left(7\gamma^2+ 8\right) -4\rho/\rho^{\scriptscriptstyle{\mathrm{I}}}_c+\left(5\gamma^2 +8\right)\sqrt{1-\rho/\rho^{\scriptscriptstyle{\mathrm{I}}}_c}}{(\gamma^2 +1)\left(1+\sqrt{1-\rho/\rho^I_c}\right)^2}\right]\frac{\rho}{\rho^{\scriptscriptstyle{\mathrm{I}}}_c}\nb\\
  &&  + 4\pi G P \left[\frac{3\gamma^2+2+2\sqrt{1-\rho/\rho^{\scriptscriptstyle{\mathrm{I}}}_c}}{(\gamma^2+1)\left(1+\sqrt{1-\rho/\rho^{\scriptscriptstyle{\mathrm{I}}}_c}\right)}\right]\frac{\rho}{\rho^{\scriptscriptstyle{\mathrm{I}}}_c},
\end{eqnarray}
where
\begin{equation}\label{key}
\rho^{\scriptscriptstyle{\mathrm{I}}}_c=\frac{\rho_{c}}{4(\gamma^2+1)} ~.
\end{equation}
We note from above equations that due to boundedness of the energy density, the Hubble rate is also generically bounded. However, the second time derivative of the scale factor $ \ddot a $ can still diverge if the pressure diverges. Hence the Ricci scalar given by $ R = 6(H^2 + \ddot a/a) $ can also diverge due to pressure divergences. We further have,
\begin{equation}\label{key}
\dot H = \frac{4\pi G (\rho +P)}{\gamma^2 +1}  \bigg(2\gamma^2 +2\frac{\rho}{\rho^{\scriptscriptstyle{\mathrm{I}}}_c}  -3\gamma^2 \sqrt{1-\frac{\rho}{\rho^{\scriptscriptstyle{\mathrm{I}}}_c}} -1\bigg),
\end{equation}
which can also diverge at pressure divergences. The above equation along with the equation of state can be used to calculate the time derivative of the Ricci scalar using $ \dot R = 6(\ddot H + 4H \dot H) $. Just like in standard LQC, $ \dot{R} $ can diverge at finite values of the energy density, Hubble rate and Ricci scalar if the time derivative of pressure $ \dot{P} $ diverges at finite pressure. Even though the details of phase space dynamics and the bounce are quite different from standard LQC, we will see in next section that the analysis of geodesics and strength of singularities in this case follows quite analogously to standard LQC. Hence, we will find that all these pressure divergences are harmless as the geodesics can be extended beyond them and they do not lead to strong singularities.

\subsubsection{The $b_{+}$ branch}

The corresponding expressions for Friedmann and Raychaudhuri equations for this root are:
\begin{eqnarray}
H^2 &=& \frac{8\pi G \alpha \rho_{\Lambda}}{3} \bigg(1-\frac{\rho}{\rc}\bigg) \bigg[1+\bigg(\frac{1-2 \gamma^2 + \sqrt{1-\rho / \rc}}{4\gamma^2 (1+\sqrt{1- \rho /\rc})}\bigg)\frac{\rho}{\rc}\bigg] , \\	
\frac{\ddot a}{a} &=& - \frac{4\pi \alpha G}{3} (\rho + 3P-2\rho_{\Lambda}) + 4\pi \alpha GP \bigg[\frac{2-3\gamma^2 +2 \sqrt{1-\rho /\rc}}{(1-5\gamma^2)(1+\sqrt{1-\rho / \rc})}\bigg] \frac{\rho}{\rc}\\
&& -  \frac{4\pi \alpha G}{3} \rho \bigg[\frac{2\gamma^2 + 5\gamma^2 (1+\sqrt{1-\rho /\rc})-4(1+\sqrt{1-\rho /\rc})^2}{(1-5\gamma^2)(1+\sqrt{1-\rho / \rc})^2}\bigg] \frac{\rho}{\rc},
\end{eqnarray}
where $ \alpha \equiv (1-5\gamma^2)/(\gamma^2 +1)  $ and $ \rho_{\Lambda} \equiv 3/[8\pi G \alpha \lambda^2 (1+ \gamma^2)^2]  $ \cite{Li2018}. Again we note that the Hubble rate is generically bounded, but the second derivative of scale factor, the Ricci scalar, and the time derivative of the Hubble rate,
\begin{equation}\label{key}
\dot H = \frac{4\pi G (\rho +P)}{\gamma^2 +1} \bigg(2\gamma^2 +2\frac{\rho}{\rc} + 3\gamma^2 \sqrt{1-\frac{\rho}{\rc}} -1\bigg),
\end{equation}
can diverge if the pressure $ P $ diverges.  We note that $ \dot{R} $ can diverge if the time derivative of pressure $ \dot{P} $ diverges even at a finite pressure so that the Ricci scalar and the time derivative of the Hubble rate are also bounded. We will show in the next section that no strong singularities exist in the effective spacetime and the geodesic are well defined at such events. Hence all such curvature pathologies will turn out to be harmless.

\subsection{Effective dynamics of mLQC-II}

While mLQC-II also quantizes the Lorentzian term in the Hamiltonian constraint independently of the Euclidean term in contrast to standard LQC, it departs from mLQC-I in the treatment of the Lorentzian term. It uses the fact that the extrinsic curvature in the spatially flat FLRW model is proportional to the Ashtekar-Barbero connection. Thus unlike mLQC-I, the Lorentzian term in mLQC-II is first re-expressed in terms of the Ashtekar-Barbero connection before using holonomies to quantize it \cite{Yang2009}. This results in a quantization which is qualitatively different from mLQC-I, leading to effective dynamics which surprisingly bears similarity with standard LQC. As in mLQC-I, the Hamiltonian constraint in mLQC-II results in a fourth order quantum difference equation leading to two branches of solutions in the effective phase space \cite{von-Neumann}. However, one of the branches turns out to be unphysical and only one branch is needed to describe the whole phase space evolution \cite{Li2018b}. Even though the modified Friedmann and Raychaudhuri equations in case of mLQC-II are more complicated than standard LQC, the bounce turns out to be symmetric with a similar pre-bounce physics as standard LQC \cite{Yang2009,Li2018b}. The effective Hamiltonian in mLQC-II is \cite{Yang2009,Li2018b},
\begin{equation}\label{key}
\mathcal{H} = - \frac{3v}{2\pi G \lambda^2 \gamma^2} \sin^2 \bigg(\frac{\lambda b}{2}\bigg) \left\lbrace 1 + \gamma^2 \sin^2 \bigg(\frac{\lambda b}{2}\bigg) \right \rbrace + \mathcal{H_\mathrm{m}},
\end{equation}
which results in the following equations of motion,
\begin{eqnarray}\label{bmlqc-II}
\dot b &=& -\frac{6\sin^2 (\lambda b/2)}{\gamma \lambda^2}\bigg(1 + \gamma^2 \sin^2 \bigg(\frac{\lambda b}{2}\bigg)\bigg) - 4\pi G \gamma P, \label{Eom_sy_modLQC1} \\
\dot v &=& \frac{3\sin(\lambda b)v}{\gamma \lambda} \bigg(1 +\gamma^2 - \gamma^2 \cos(\lambda b)\bigg). \label{Eom_sy_modLQC2}
\end{eqnarray}

Integrating the volume equation we get,
\begin{equation}\label{key}
v(t)= v_o \exp \left\lbrace \int^t_{t_o} \mathrm{dt} \frac{3\sin(\lambda b)v}{\gamma \lambda} \bigg(1 +\gamma^2 - \gamma^2 \cos(\lambda b)\bigg) \right\rbrace.
\end{equation}
Since the integrand inside the exponential is bounded, the volume will remain non-zero and finite for all finite time evolution. Analogous to standard LQC and mLQC-I, the above evolution equations for phase space variables are linear first order ODEs with bounded coefficients except for the pressure term. Thus the phase space in mLQC-II is also maximally extendible except for pressure divergences.

We further look at the behavior of energy density. From the vanishing of the Hamiltonian constraint, the energy density is dynamically equal to,
\begin{equation}\label{mLQC-II-rho}
\rho = \frac{3}{2\pi G \lambda^2 \gamma^2} \sin^2 \bigg(\frac{\lambda b}{2}\bigg) \left\lbrace 1 + \gamma^2 \sin^2 \bigg(\frac{\lambda b}{2}\bigg) \right \rbrace.
\end{equation}
Thus, the energy density is generically bounded in this case as well. Just like mLQC-I, the energy density equation is a quadratic equation for $ \sin^2(\lambda b/2) $ and leads to two solutions for $ b $. However, as shown in \cite{Li2018b}, one of the roots leads to imaginary values of the phase space variable $ b $, and the remaining real root spans the entire phase space trajectory obtained by the Hamilton's equations \eqref{Eom_sy_modLQC1} and \eqref{Eom_sy_modLQC2}. This is in contrast to mLQC-I where we needed both the solutions to span the whole phase space trajectory \cite{Li2018}. Here we discuss only the dynamics obtained from the real-valued root of equation \eqref{mLQC-II-rho} which corresponds to the physical dynamics obtained from effective Hamilton's equations. The modified Friedmann and Raychaudhuri equations are given as follows,
\begin{eqnarray}
H^2 &=& \frac{8\pi G}{3} \rho \bigg(1+\gamma^2 \frac{\rho}{\rho_{c}}\bigg) \bigg[1-\frac{(\gamma^2 +1)\rho / \rho_{c}}{(1+ \sqrt{1+\gamma^2\rho / \rho_{c}})^2} \bigg] , \\
\frac{\ddot{a}}{a} &=& -\frac{4\pi G}{3}(\rho+3P)-4\pi GP \bigg[\frac{3\gamma^2 +1-2\sqrt{1+\gamma^2 \rho /\rho_{c}}}{1+\sqrt{1+\gamma^2 \rho/\rho_{c}}}\bigg]\frac{\rho}{\rho_{c}} \\
&& -\frac{4\pi G \rho}{3}\bigg[\frac{7\gamma^2-4\gamma^2 \rho/\rho_{c}-1+(5\gamma^2-3)\sqrt{1+\gamma^2 \rho/\rho_{c}}}{(1+\sqrt{1+\gamma^2 \rho/\rho_{c}})^2}\bigg]\frac{\rho}{\rho_{c}}.
\end{eqnarray}

It turns out that these equations lead to a symmetric bounce qualitatively similar to that obtained in standard LQC \cite{Li2018b}. The critical energy density $ \rho^{\scriptscriptstyle{\mathrm{II}}}_{c} $ (where superscript II corresponds to mLQC-II) at the bounce can be obtained from the above equations by using $ H=0 $ and $ \ddot{a} > 0 $, which gives,
\begin{equation}\label{key}
\rho^{\scriptscriptstyle{\mathrm{II}}}_{c}=4(\gamma^2 +1)\rho_{c}.
\end{equation}

Since the energy density is generically bounded, the Hubble rate is also generically bounded. As in standard LQC and mLQC-I, the second derivative of the scale factor $ \ddot{a} $ and consequently the Ricci scalar $ R $ can diverge if the pressure diverges. The time derivative of the Hubble rate $ \dot H $ turns out to be 
\begin{equation}\label{key}
\dot{H}=\frac{4\pi G(\rho +P)}{3}\bigg[3+2\gamma^2(1+\rho/\rho_{c})-3(\gamma^2 +1)\sqrt{1+\gamma^2 \rho/\rho_{c}}\bigg],
\end{equation}
which can also diverge due to divergence in pressure. From the time derivative of Ricci scalar we find that $ \dot{R} $ can diverge even when the Ricci scalar and the time derivative of the Hubble rate are finite, if $ \dot{P} $ diverges at finite pressure. But as in case of standard LQC and mLQC-I, these turn out to be weak singularities beyond which geodesics can be extended as shown in the following section. 

\section{Geodesic Completeness and Strength of Singularities}

As discussed in Sec. I, spacetime singularities can be characterized by geodesic incompleteness and their strength which captures properties of the tidal forces at the singular event. Notably not all curvature pathologies are guaranteed to be true singularities if geodesics can be extended beyond them and singularities turn out to be weak, examples of which are type-II and type-IV singularities discussed in the next section. On the other hand, strong singularities are true singularities with respect to geodesic incompleteness. In the following, we will use features noted in Sec. II to understand geodesic completeness and strength of singularities in modified isotropic loop cosmologies.  It turns out that as in standard LQC, effective spacetime is geodesically complete in mLQC-I and mLQC-II, and there exist no strong singularities in these loop cosmologies. Our treatment in this section will be general without assuming any particular equation of state of matter. Phenomenological examples will be discussed in the next section. Even though these studies have been carried out earlier for standard LQC \cite{ps09}, for comparative completeness we will discuss results in standard LQC below.

In the previous section, we showed that the quantum geometry effects ensure the Hubble rate (and hence the expansion scalar), and the energy density  are generically bounded and the volume remains non-zero and finite in LQC effective dynamics for the standard LQC as well as for mLQC-I and mLQC-II. This hints that many singular events that are associated with the divergence in these quantities may be absent in these loop cosmologies. But as noted earlier, interestingly some curvature invariants such as the derivative of the Hubble rate, the Ricci scalar and its time derivative may still diverge in above loop cosmologies if the isotropic pressure or its time derivative diverges at finite values of energy density, Hubble rate and volume. Such divergences require some exotic equations of state, which will be discussed in Sec. IV. Thus, even in the presence of quantum geometric effects curvature pathologies can exist. As we will show these divergences turn out to be harmless. Geodesics are well behaved at such events and all such curvature pathologies turn out to be weak singularities. 

Let us begin with analysis of geodesics.  The geodesic equations for the FLRW spacetime in Cartesian coordinates turn out to be,
\begin{eqnarray}\label{geodesic1}
x'' &=& - \frac{a'}{a}x' , \quad y'' = - \frac{a'}{a}y' , \quad z'' = - \frac{a'}{a}z' ,\\
t'' &=& - \frac{aa'}{t'} (x'^2 +y'^2 + z'^2).
\end{eqnarray}
Here prime denotes derivative with respect to the affine parameter. The above set of equations give us the accelerations along the geodesics. These can be integrated to find the velocities along the geodesics,
\begin{eqnarray}\label{geodesic2}
x' &=& C_x / a^2, \quad y' = C_y / a^2, \quad z' = C_z / a^2, \\
t'^2 &=& \epsilon + \frac{C_x^2 + C_y^2 + C_z^2}{a^2}, 
\end{eqnarray}
where $ C_x, C_y $ and $ C_z $ are constants of integration and $\epsilon$ vanishes for null geodesics and is unity for massive particles.  These geodesic equations tell us that accelerations and velocities along the geodesics diverge at points where either the scale factor vanishes and/or the Hubble rate diverges. For example, at the big bang singularity, the scale factor vanishes and Hubble rate diverges leading to the break down of geodesic equations. However, we showed in previous section that for all three loop cosmologies, i.e.  LQC, mLQC-I and mLQC-II, the scale factor $a = v^{1/3}$ is non-zero and finite for all finite time evolution, and  the Hubble rate is generically bounded for all the time. Though the energy density is universally bounded in LQC, mLQC-I and mLQC-II, it is to be noted that the pressure is not which in principle can diverge causing curvature pathologies. Notably, even for such pathologies which include divergences of Ricci scalar and/or its derivative, geodesic equations do not break down in the effective spacetime description of LQC, mLQC-I and mLQC-II since the Hubble rate remains bounded and the scale factor is non-zero and finite at such events. In this sense the effective spacetime is geodesically complete and curvature pathologies do not result in breakdown of geodesics. 

We now analyze the strength of potential singularities in modified loop cosmologies. A strong singularity is defined to be one which crushes all in-falling objects to zero volume irrespective of their properties. A singularity which does not satisfy this criterion is referred to as a weak singularity and can be considered to be harmless since, by definition, a sufficiently strong object may survive such an event. The necessary and sufficient conditions for a potential singularity to be a strong singularity were derived by Tipler \cite{tipler1977singularities}, and Kr{\'o}lak \cite{krolak,ck-1985conditions}. In the following we consider the conditions provided by Kr{\'o}lak which are broader than those provided by Tipler.

If a given singularity, located at $ \tau_o $ along a non-spacelike geodesic running into the singularity, is a strong singularity then the following integral evaluated along the geodesic parameterized by affine parameter $\tau$ 
\begin{equation}
K^i_j = \int^{\tau}_0 \mathrm{d}\tau |R^i_{~0j0}| \label{krolak_condition}
\end{equation}
must diverge as $ \tau \rightarrow \tau_o $, where the components of the Riemann tensor are evaluated in a parallel-propagated frame along the geodesic. Note that this is a necessary but not sufficient condition for a strong singularity. Further, note that in a homogeneous and isotropic universe like the one under consideration, the only timelike geodesics we need to consider are the matter world lines followed by fundamental observers for which the proper time along the geodesics is $ t $ itself. Further, in case of null geodesics we can take the affine parameter equal to time $t$ itself. %So the above integral can be considered to be evaluated over $ t $.

For the spatially flat FLRW spacetime, the non-vanishing components of the Riemann tensor are:
\begin{eqnarray}
R^0_{~110} &=& R^0_{~220} = R^0_{~330} = -a \ddot a, \\
R^1_{~010} &=& R^2_{~020} = R^3_{~030} = - {\ddot a}/a = -H^2 - \dot H, \\
R^1_{~221} &=& R^1_{~331} = R^2_{~332} = -R^2_{~121} = -R^3_{~131} = -R^3_{~232} = - {\dot a}^2.
\end{eqnarray}
%The above mentioned condition \eqref{krolak_condition} is necessary condition for a strong singularity which involves an integral of the Riemann tensor components. 
In LQC models the Hubble rate $ H={\dot a}/a $ is bounded and both $ \dot a $ and $ a $ remain finite and non-zero for all finite times, and any potential divergence in Riemann tensor components only comes through the $ \ddot a $ terms. Hence, even for potential curvature pathologies the above mentioned integral of Riemann tensor components does not diverge as all the second derivatives of the scale factor are removed after one integration. Such a curvature pathology is a weak singularity in LQC, mLQC-I and mLQC-II. Thus, fundamental observers following matter world lines see no strong singularities in spatially flat isotropic LQC, mLQC-I and mLQC-II in finite proper time evolution.  One can in principle choose a time coordinate not corresponding to fundamental observers following matter world lines, say $\tilde t$ related to $t$ as $\tilde N \d \tilde t = \d t$,  such that $ \tilde t $ is finite when the proper time $t$ for fundamental observers becomes infinite, e.g. when $\d \tilde t = e^{-t} \d t $. Thus a potential strong singularity may occur in a finite $\tilde t$ evolution, however the above result implies that fundamental observers see such an event only in infinite time $t$ evolution (for further discussion see Sec. V of Ref. \cite{ks-strong}).

In contrast to the above result that strong singularities are absent in finite proper time evolution in loop cosmological models considered here, curvature pathologies in the classical theory can be strong as well as weak singularities. As an example, we demonstrate that the big bang singularity in classical spacetime is a strong singularity. Since the Hubble rates diverge at the big bang singularity in the classical dynamics, one can see that the necessary condition \eqref{krolak_condition} is satisfied. However, to show that it is indeed a strong singularity, we need to consider the sufficient conditions for a potential curvature pathology to be a strong singularity. As discussed in \cite{ck-1985conditions}, a sufficient condition for a singularity to be a strong curvature singularity can be given as follows:

For a non-spacelike geodesic parameterized by an affine parameter $ \tau $, if the integral
\begin{equation}\label{suff_condition}
I = \int^{\tau}_0 |R_{00}| \mathrm{d}\tau
\end{equation}
diverges in a finite distance $ \tau $ along the geodesic, then the singularity is a strong singularity. From the values of the Riemann tensor components given above, we have $ R_{00} = -3 \ddot{a}/a $. Using the classical Raychaudhuri equation \eqref{Raychaudhari_FRW_classical}, the integral \eqref{suff_condition} for comoving observers becomes,
\begin{equation}\label{key}
I= 3H(t) + 3 \int^t_0 \mathrm{d}t' H^2(t').
\end{equation}
We see that the big bang singularity in the classical spacetime is a strong singularity as the Hubble rate diverges at the big bang singularity, leading to the divergence of the integral \eqref{suff_condition}.

Thus, we conclude that unlike the classical dynamics of the FLRW model, no strong singularities in the effective loop cosmological spacetimes are encountered in a finite time evolution owing to the boundedness of Hubble rate and non-vanishing of the scale factor in finite time. There can be pressure divergences which result into unbounded values of curvature invariants but these amount to weak singularities since these divergences do not satisfy the necessary conditions for strong singularities. This corroborates our conclusion from analysis of geodesics that these pressure divergences are essentially harmless. In the next section, we illustrate the nature of singularity resolution in loop cosmological models by numerically finding solutions of effective spacetimes in loop cosmologies and comparing them with the classical spacetime and amongst themselves for a phenomenological equation of state allowing various exotic singularities.

\section{Phenomenological modeling of singularities}

In this section we illustrate results regarding the resolution of singularities in loop cosmological models obtained in previous sections using a phenomenological description for equation of state which permits all known singularities in the classical FLRW spacetime. The matter model we study was first discussed in Ref. \cite{Nojiri2005}, which has been used earlier in LQC to study singularity resolution \cite{ps09,psfv}. Our goal is to compare solutions of GR and LQC with mLQC-I and mLQC-II by numerically solving dynamical equations in Sec. II. We are interested in four types of singularities which are distinct from the conventional big bang/crunch singularities studied earlier \cite{Li2018,Li2018b}. These are: (i) type-I singularities or big rip singularities where energy density and Hubble rate diverge at an infinite value of scale factor in finite time evolution, (see for e.g. Refs. \cite{bigrip}), (ii) type-II or sudden singularities where energy density and Hubble rate are finite, but pressure diverges at a finite value of scale factor in finite time \cite{sudden}, (iii) type-III or big freeze singularities where energy density and Hubble rate diverge at a finite value of scale factor \cite{Nojiri2005,freeze}, and (iv) type-IV singularities where energy density, pressure, Hubble rate and Ricci scalar are finite but time derivative of Ricci scalar diverges \cite{Nojiri2005,type4}.

In the phenomenological model for the equation of state which exhibits all the cosmological singularities of interest, we consider the following ansatz \cite{Nojiri2005}:
\begin{equation}\label{key}
P = -\rho - f(\rho), ~~~\mathrm{where} ~~~ f(\rho) = \frac{AB \rho^{2\alpha-1}}{A \rho^{\alpha-1}+B}.
\end{equation}
Here $ A,B $ and $ \alpha $ are parameters which can be varied to generate different types of singularities.
Using the matter-energy conservation equation, $\dot \rho = -3H(\rho +P)$ 
which remains valid in all the different effective models of loop cosmology, we can find the scale factor as a function of the energy density:
\begin{equation}\label{key}
a=a_o \exp \left \lbrace \frac{1}{6} \frac{(2A+B\rho^{(1-\alpha)})\rho^{(1-\alpha)}}{AB(1-\alpha)}\right\rbrace .
\end{equation}
This equation can be inverted to give,
\begin{equation}\label{key}
\rho = \bigg(-\frac{A}{B}\pm \bigg(\frac{A^2}{B^2}-6A(\alpha-1)\ln\bigg(\frac{a}{a_o}\bigg) \bigg)^{1/2}\bigg)^{1/(1-\alpha)}.
\end{equation}
The time derivative of the pressure then turns out to be 
\begin{equation}
\dot P = - \dot \rho (1+f'(\rho)) = 3H(\rho + P)(1+f'(\rho)).	  \label{dotP}
\end{equation}
Together with the equations we have derived for different models in Sec. II, we now have all the relevant equations to numerically analyze different types of singularities in the classical and various loop cosmological spacetimes. In the following subsections, we  consider each of the above four types of singularities and analyze and contrast results of our simulations. To perform these simulations we will fix the value of Barbero-Immirzi parameter to $\gamma \approx 0.2375$ using black hole thermodynamics \cite{meissner}, as is conventionally done in LQC.

\subsection{Type-I Singularities}
\begin{figure}
	\centering
	\includegraphics[width=0.49\textwidth]{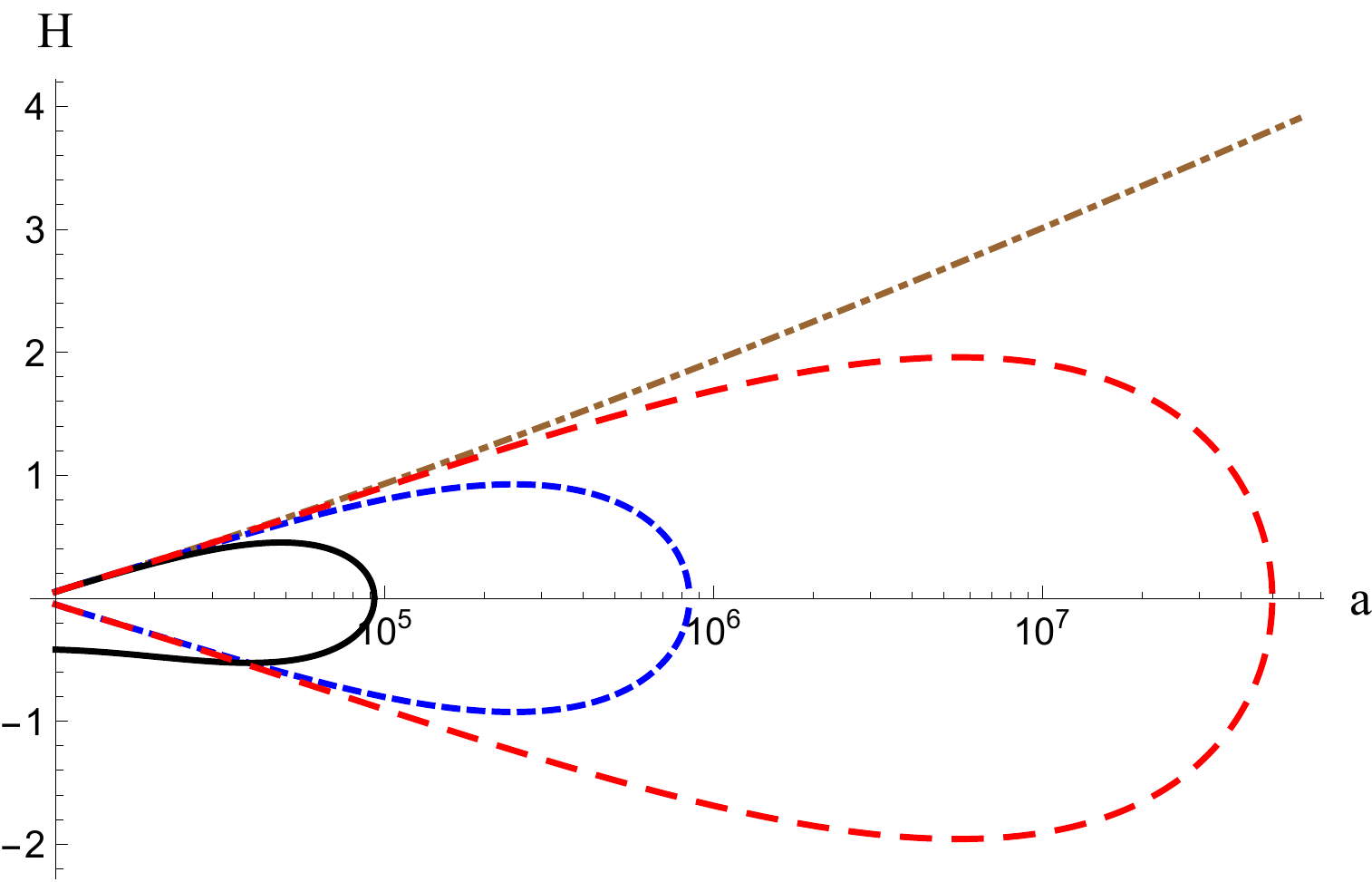} 
	\includegraphics[width=0.49\textwidth]{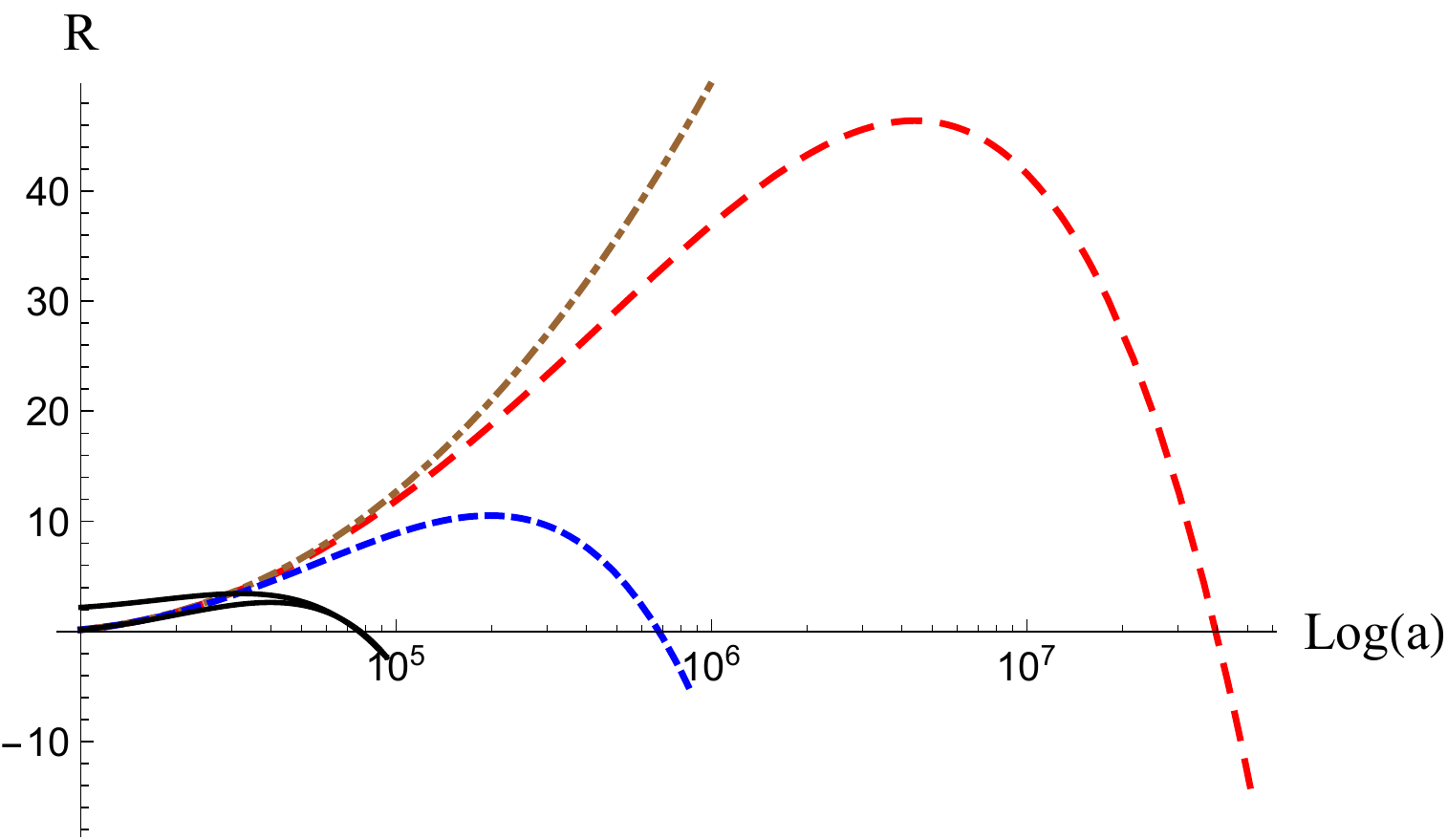} 
	\caption{In these graphs, we have plotted the curves for the Hubble rate and Ricci scalar as a function of the scale factor for the classical dynamics (dot-dashed), standard effective LQC dynamics (dashed), mLQC-I (solid) and mLQC-II (large-dashed). The values of the parameters chosen are $ A=0.1, B=-1, a_o=10000 $ and $ \alpha=0.8 $.}
	\label{fig:typeI}
\end{figure}

In eq. (\ref{key}), if we choose $ 3/4 < \alpha < 1 $ and $ A $ to be positive, we get a big rip (type-I) singularity in classical GR \cite{Nojiri2005}. This type of singularity is characterized by the divergence of the scale factor, energy density and pressure in a finite time evolution, and is an example of a strong singularity \cite{Jambrina1,ps09}. Note that for these values of parameters classically there is no initial singularity in this scenario. In order to compare the dynamics of different effective theories with the classical behavior, we plot the respective Hubble rates and the Ricci scalars in Fig. \ref{fig:typeI}. 
As expected, the Hubble rate and the Ricci scalar diverge for classical GR which is depicted by a dot-dashed curve.  However, all the curves corresponding to three loop cosmological models are very different from the classical trajectory at large values of scale factor where the classical big rip singularity occurs. This is due to the maximum value of energy density  in all the three loop cosmological models at which the Hubble rate vanishes and $ \ddot a $ switches its sign and becomes negative. This essentially means that the big rip singularity is absent in all of the loop cosmological models considered here and is replaced by a quantum recollapse at late times. The quantum recollapse is analogous to the quantum bounce often found in loop cosmological models for equations of state of matter satisfying weak energy condition. Hence, all loop cosmology curves  turn around to small values in the first plot.  Resolution of the big rip singularity and existence of quantum recollapse in isotropic LQC was first seen in Refs. \cite{sst,ps09}, and now our analysis confirms the same for mLQC-I and mLQC-II. Since the upper bound on energy density is numerically different in the three loop cosmological models, the recollapse occurs at different times, as can be seen in the plots in Fig. \ref{fig:typeI}. 

Even though most of the details of the quantum recollapse are qualitatively similar in all the loop cosmological models, there is one prominent feature that stands out. Namely, the recollapse is asymmetric in the case of mLQC-I while it is symmetric in LQC and mLQC-II.  The reason for this is tied to the properties of modified dynamical equations in mLQC-I \cite{Li2018} which were discussed in Sec. II. Specifically, solutions in the phase space consist of two different patches, which in the present case are joined at the quantum turnaround which in this case is a quantum recollapse.  Further, solutions of LQC, mLQC-I and mLQC-II agree with classical trajectory when scale factor is small, except in the $ b_{+} $ branch of mLQC-I due to the above mentioned asymmetry. This asymmetry is visible in the curve for the Hubble rate for the negative values, and in the behavior of Ricci scalar.  Further, in each of the loop cosmological models the recollapse occurs at different times because of the differences in the maximum energy density.

\subsection{Type-II Singularities}
\begin{figure}[tbh!]
	\centering
	\includegraphics[width=0.49\textwidth]{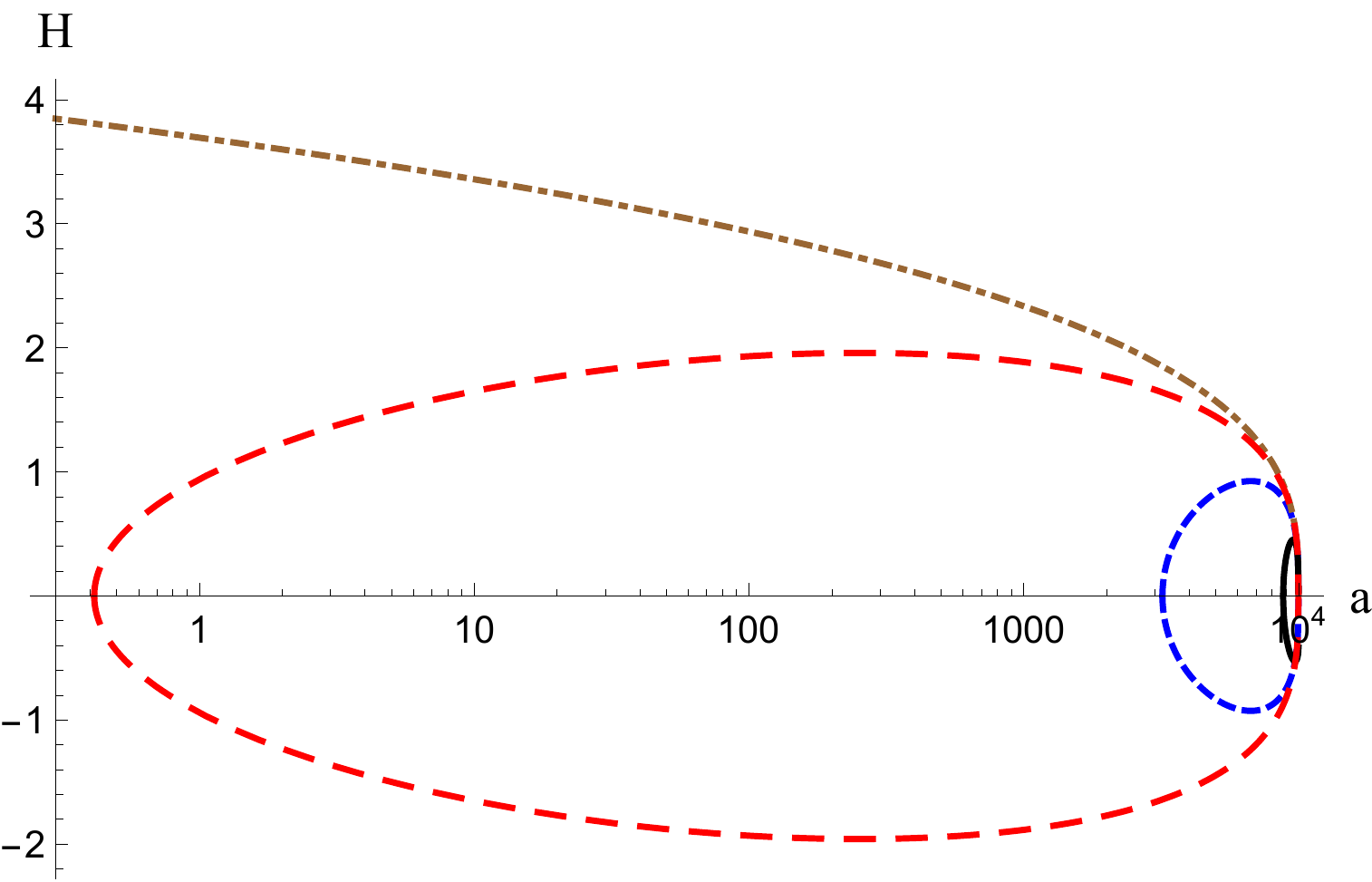} 
	\includegraphics[width=0.49\textwidth]{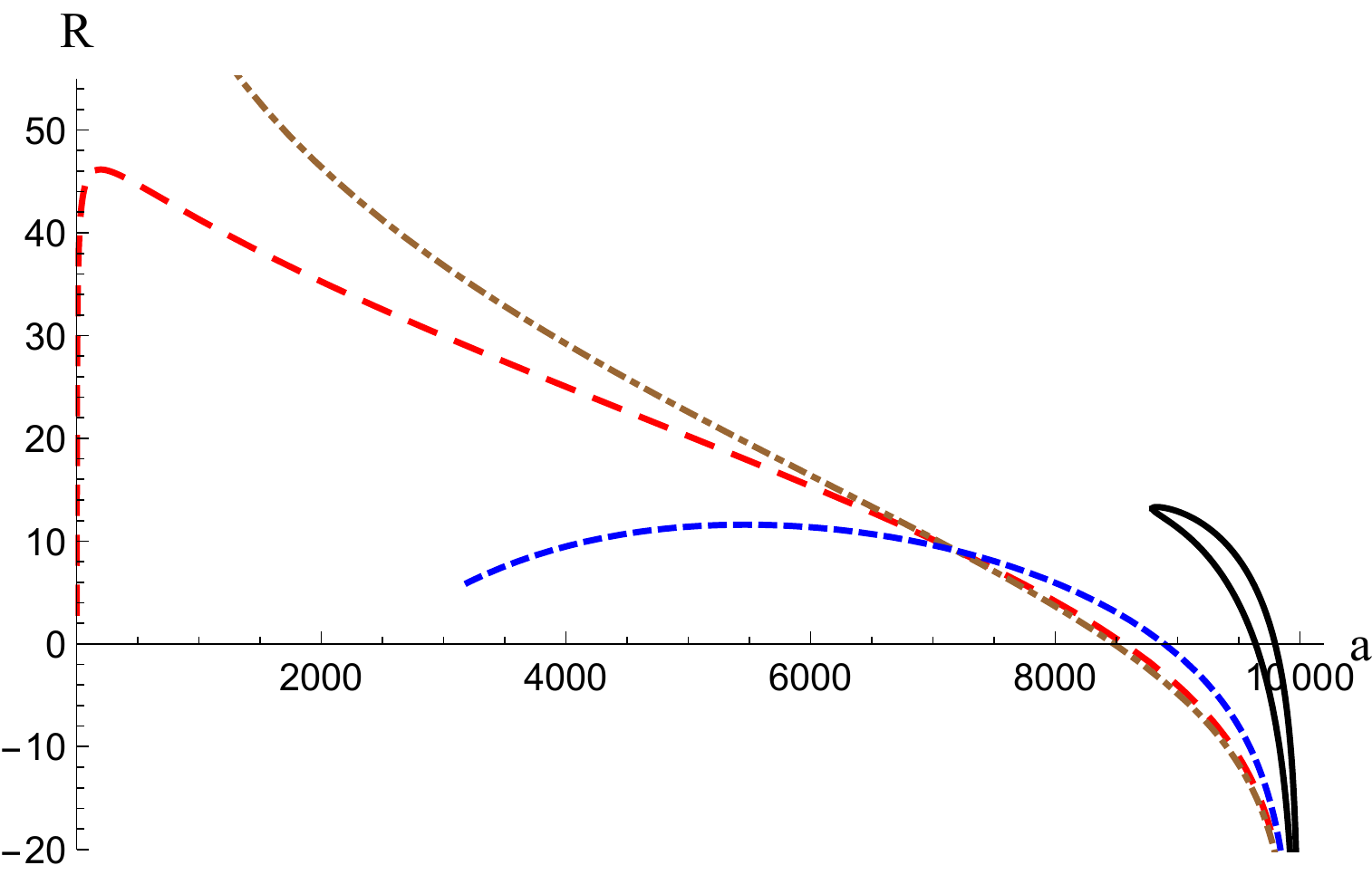} 
	\caption{In these graphs, we have plotted the curves for the Hubble rate and Ricci scalar as a function of the scale factor for the type-II singularity for the classical dynamics (dot-dashed), standard effective LQC dynamics (dashed), mLQC-I (solid) and mLQC-II (large-dashed). The values of the parameters chosen are $ A=-0.05, B=100, a_o=10000 $ and $ \alpha=1/4.1 $. Note that the weak singularity occurs at $a_o = 10000$ both in the pre-bounce and the post-bounce branches for LQC and mLQC-II. The weak singularity is present only in the post-bounce branch in mLQC-I.}
	\label{fig:typeII}
\end{figure}

\begin{figure}
	\centering
	\includegraphics[width=0.49\textwidth]{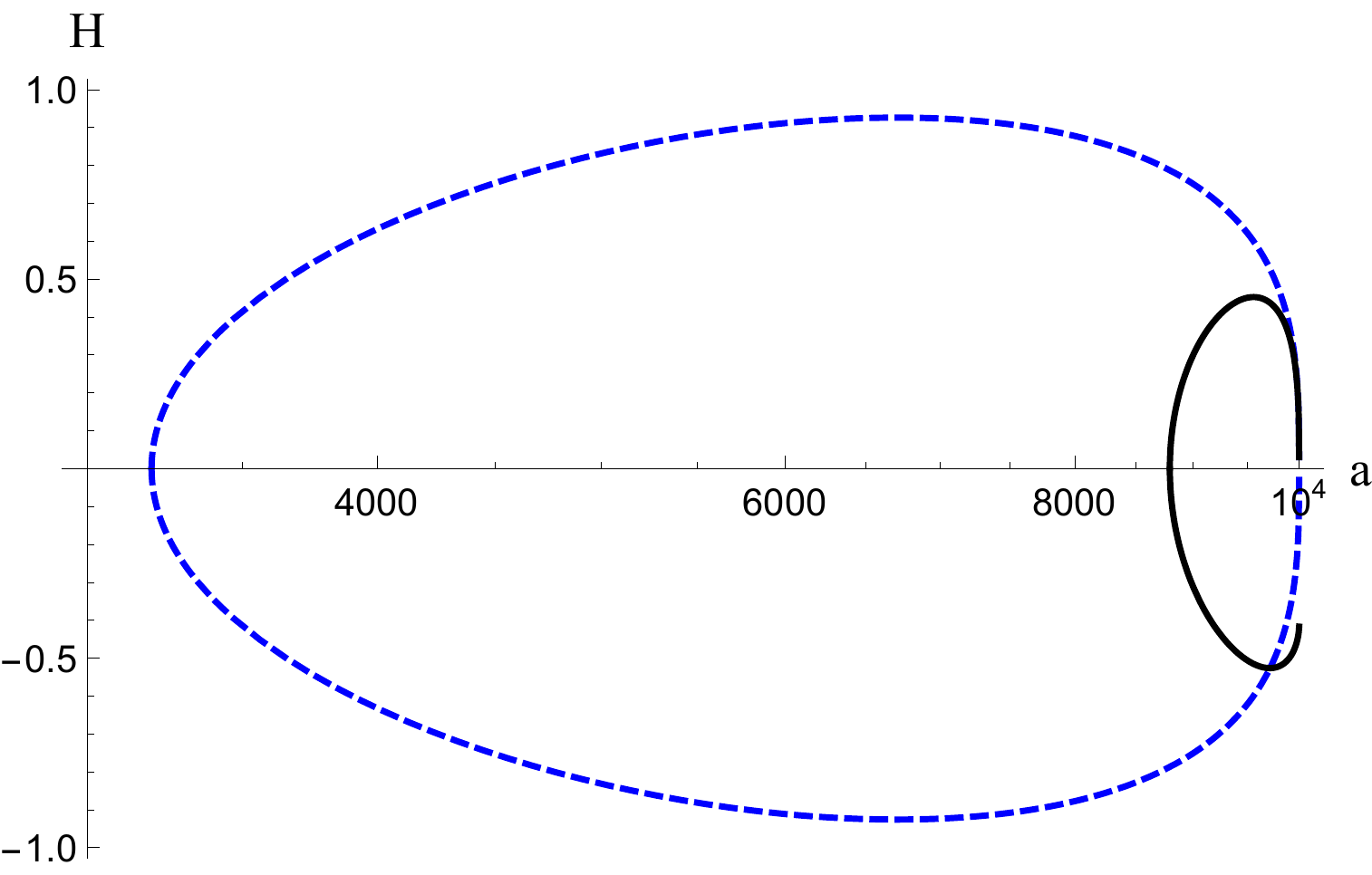} 
	\includegraphics[width=0.49\textwidth]{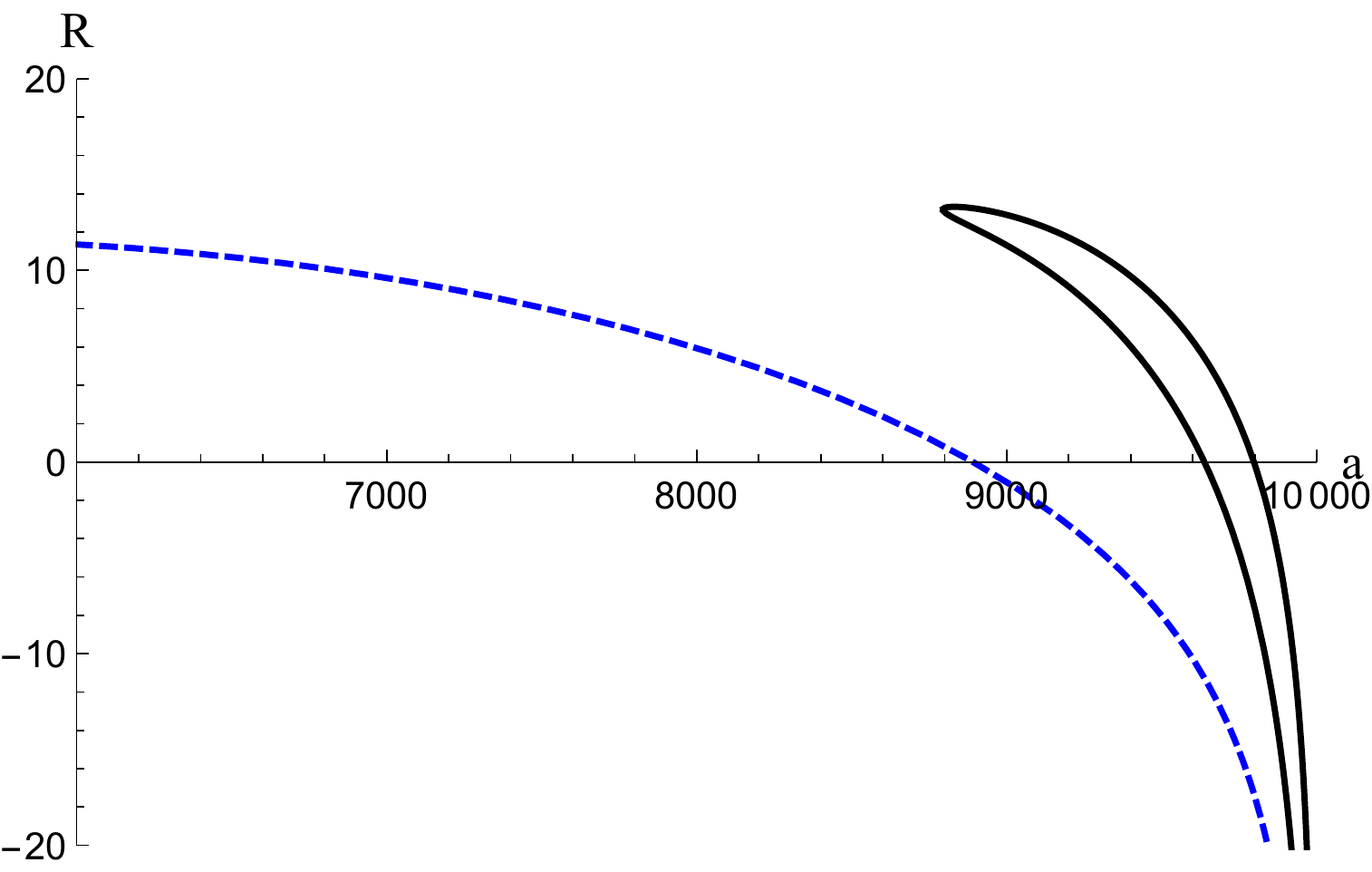} 
	\caption{These graphs contain the curves for the Hubble rate and Ricci scalar as a function of the scale factor for the type-II singularity for mLQC-I (solid) and standard LQC (dashed) for a closer comparison. The values of the parameters chosen are $ A=-0.05, B=100, a_o=10000 $ and $ \alpha=1/4.1 $.}
	\label{fig:typeIIA}
\end{figure}

A necessary condition for these sudden singularities to occur is to have $ A/B<0 $ in eq.(\ref{key}). In classical GR there is a big bang singularity in this model in the past as $ a \rightarrow 0 $. This can be seen in the behavior of Ricci scalar for the classical case in Figs. \ref{fig:typeII} and \ref{fig:typeIIA}, where we plot the Hubble rates and the Ricci scalars of different dynamical models for comparison. In type-II singularity, the energy density and the Hubble rate both remain finite and go to zero as the future  singularity is approached at $a \rightarrow a_o $, however the Ricci scalar diverges due to the divergence in pressure. Since this singularity occurs when pressure becomes infinite the evolution equation for $b$ in classical theory (\ref{bcl}), LQC (\ref{blqc}), mLQC-I (\ref{bmlqc-I}) and mLQC-II (\ref{bmlqc-II}) breaks down. 

Figs. \ref{fig:typeII} and  \ref{fig:typeIIA} demonstrate that the past  big bang singularity is replaced by big bounce where the Hubble rates vanish while Ricci scalar remains finite. We see  that in Fig. \ref{fig:typeII}, the big bang singularity of the classical thoery is resolved in all loop cosmological models considered here and is replaced by a bounce where Hubble rate vanishes. It should be noted that evolution of Hubble rate does not continue past the weak singularity at $a=a_o$, and there is no recollapse at weak singularity as Fig. \ref{fig:typeII} might suggest if one observes the behavior of Hubble rate near $a = a_o$, which is an artifact of plotting pre-bounce and post-bounce branches together. Rather, the weak singularity exists both in the pre-bounce branch and the post-bounce branch at $a=a_o$ for LQC and mLQC-II, a feature shared for type-IV singularity. That there is no evolution of Hubble rate across weak singularity  can be seen in Fig. \ref{fig:typeIIA}. One can clearly see that Hubble rate evolves continuously across the bounce, but approaches zero independently in pre-bounce and post-bounce branches at $a = a_o$.  

It is important to note that the type-II singularity is still not removed by loop cosmological effective dynamics in any of the cases considered. This is because the energy density remains very low compared to the critical density when  type-II singularities occur not allowing quantum gravity effects to become significant. Due to the exotic equation of state, pressure diverges at a finite energy density and a finite Hubble rate, causing divergence in the Ricci scalar  as $ a \rightarrow a_o $. In the limit when $ a \rightarrow a_o $, loop cosmological dynamics in all three cases closely approximates the classical dynamics since energy density is much smaller than maximum energy densities in LQC, mLQC-I and mLQC-II, as a result of which quantum geometric modifications to Friedmann and Raychaudhuri equations are negligible. It is to be noted that the above curvature pathology is not a strong singularity. Since the Hubble rate is bounded in all loop cosmological models, we note from the analysis in the previous section that the integral (\ref{krolak_condition}) is finite in the above evolution. Hence the singularity is a weak singularity. Since the Hubble rate is bounded and the type-II singularity occurs at a finite value of scale factor, we find that geodesic equations (3.1--3.4) do not break down at this singularity. Since the time derivative of phase space variable $b$ blows up at this singularity, $b$ is indeterminate. As a result, even though the volume and its time derivative, which capture the metric and the Christoffel symbols, are always finite from Hamilton's equations in LQC, mLQC-I and mLQC-II, their exact values beyond the type-II singularity can not be determined. Note that this indetreminism beyond the type-II singularity holds also for GR, as can be seen from Hamilton's equations in Sec. IIA, since the singularity occurs in the regime where quantum geometric corrections are negligible. In this sense, all these models lose some predicitability about the geodesics beyond the type-II singularity. 

%expansion scalar is bounded and singularity occurs at a finite value of scale factor, the geodesic evolution does not break down in this case. 

As expected, the time taken from the big bounce to the type-II singularity is different amongst LQC, mLQC-I and mLQC-II cases because their critical densities differ from each other. Again, the big bounce which replaces the big bang singularity is asymmetric in case of mLQC-I while it is symmetric in other two cases. The asymmetry is due to switching of the branches in case of mLQC-I for the same reasons as mentioned before. Notably, LQC and mLQC-II have a type-II singularity in the past pre-bounce branch as well where the Ricci scalar diverges while the Hubble rates vanish except in mLQC-I case where the Hubble rate remains at a finite and non-zero value.

\subsection{Type-III Singularities}

For $ \alpha>1 $, the phenomenological model under consideration leads to a type-III or the big freeze singularity when $ a \rightarrow a_o $ at a finite time. This is a strong singularity in classical GR accompanied with divergences in energy density, pressure and the Hubble rate. As in the case of type-I singularity, there is no initial singularity in this model classically.

\begin{figure}[tbh!]
	\centering
	\includegraphics[width=0.49\textwidth]{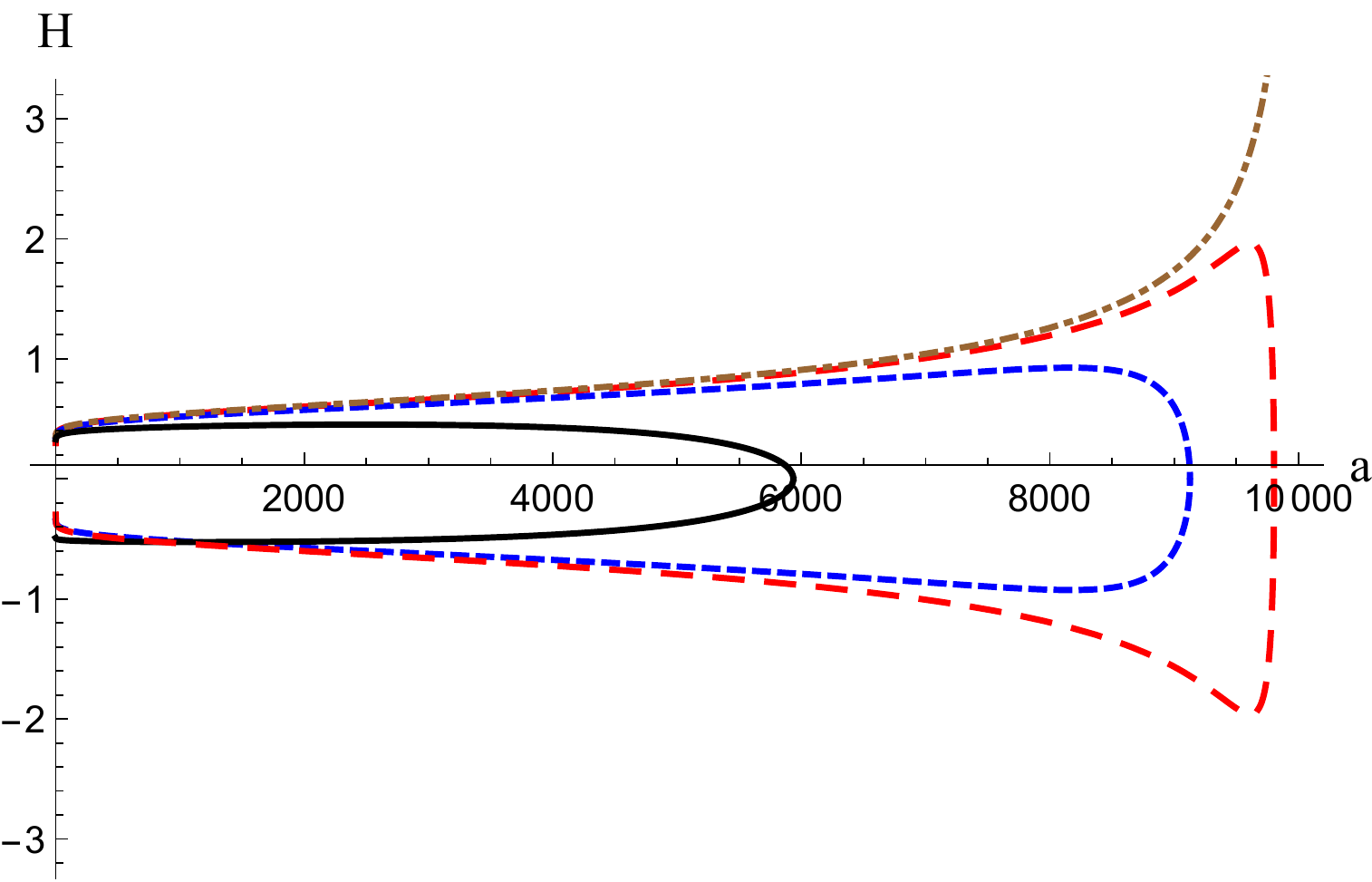} 
	\includegraphics[width=0.49\textwidth]{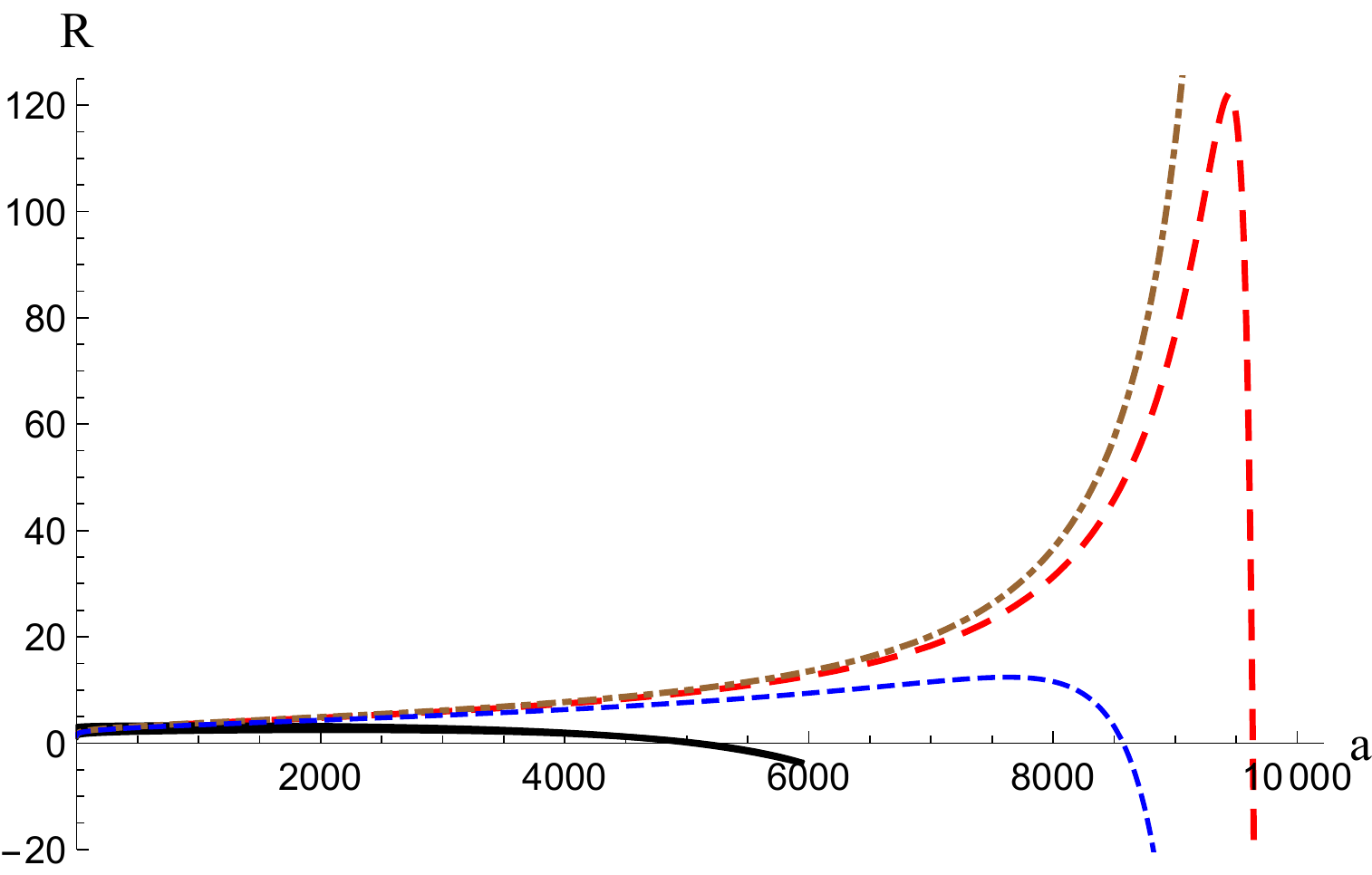} 
	\caption{In these graphs, we have plotted the curves for the Hubble rate and Ricci scalar for the type-III singularity as a function of the scale factor for the classical dynamics (dot-dashed), standard effective LQC dynamics (dashed), mLQC-I (solid) and mLQC-II (large-dashed). The values of the parameters chosen are $ A=100, B=10, a_o=10000 $ and $ \alpha=2 $.}
	\label{fig:typeIII}
\end{figure} 

A comparison of the various dynamical models is plotted in Fig. \ref{fig:typeIII}. We find that the loop cosmological models agree very closely with the classical curve at small values of scale factor when the energy density is small. But they depart from the classical trajectory as the classical future singularity is approached and the energy density becomes comparable to the maximum density in modified Friedmann dynamics. Since the Hubble rate in each of the loop cosmological models vanish when the corresponding maximum energy density is reached, the quantum turnaround results in classical singularity replaced by a quantum recollapse. We find that the Ricci scalar remain bounded in all loop cosmological models and retrace their values after the recollapse except in the case of mLQC-I where the recollapse is asymmetric.

Finally, there are expected qualitative differences amongst LQC, mLQC-I and mLQC-II, namely that the recollapse occurs at different times in them because their critical densities are different. And, dynamics of mLQC-II gives an asymmetric recollapse where the evolution after the recollapse follows a trajectory different from the one before the recollapse.

\subsection{Type-IV Singularities}

\begin{figure}[tbh!]
	\centering
	\includegraphics[width=0.49\textwidth]{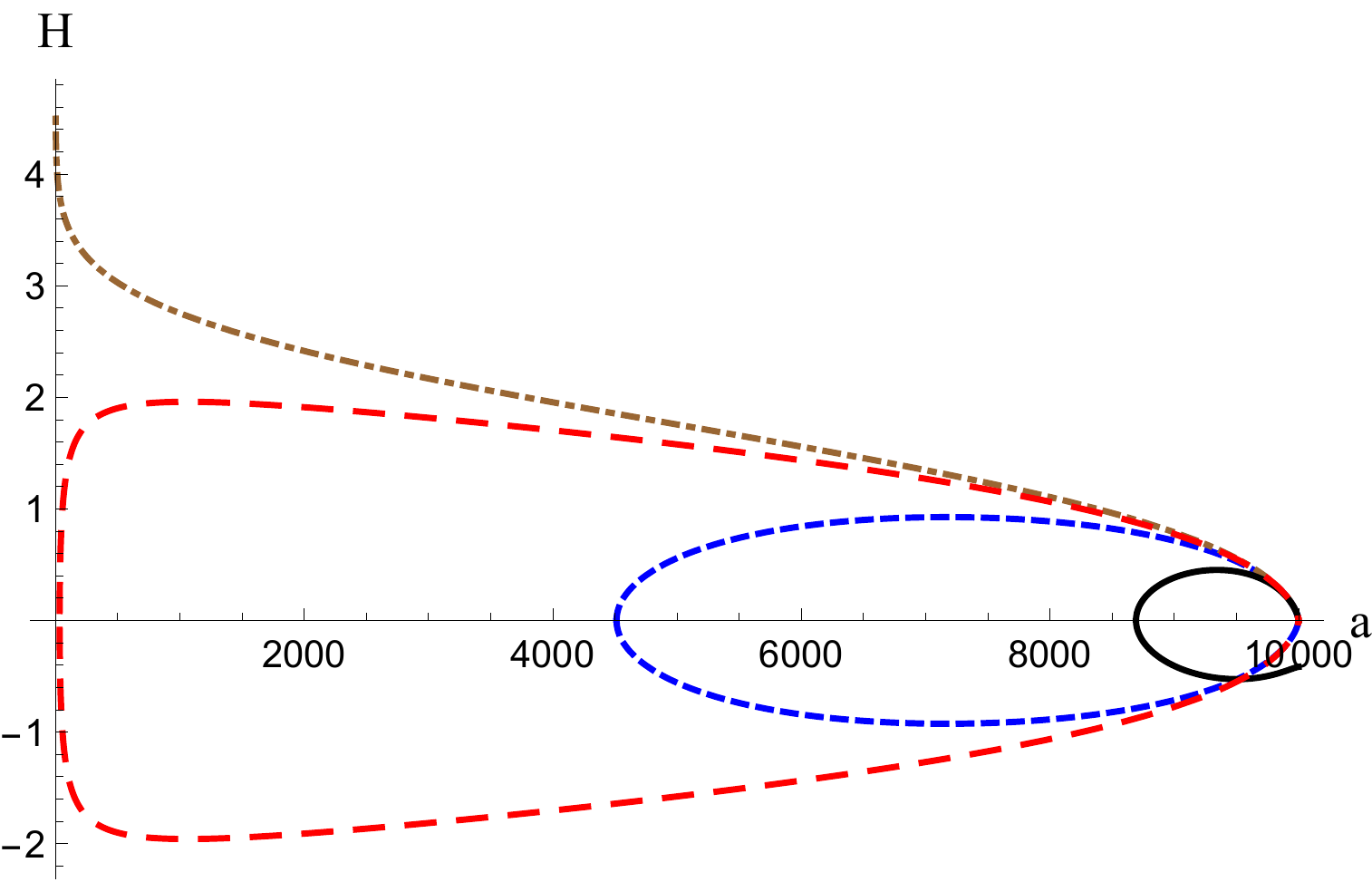} 
	\includegraphics[width=0.49\textwidth]{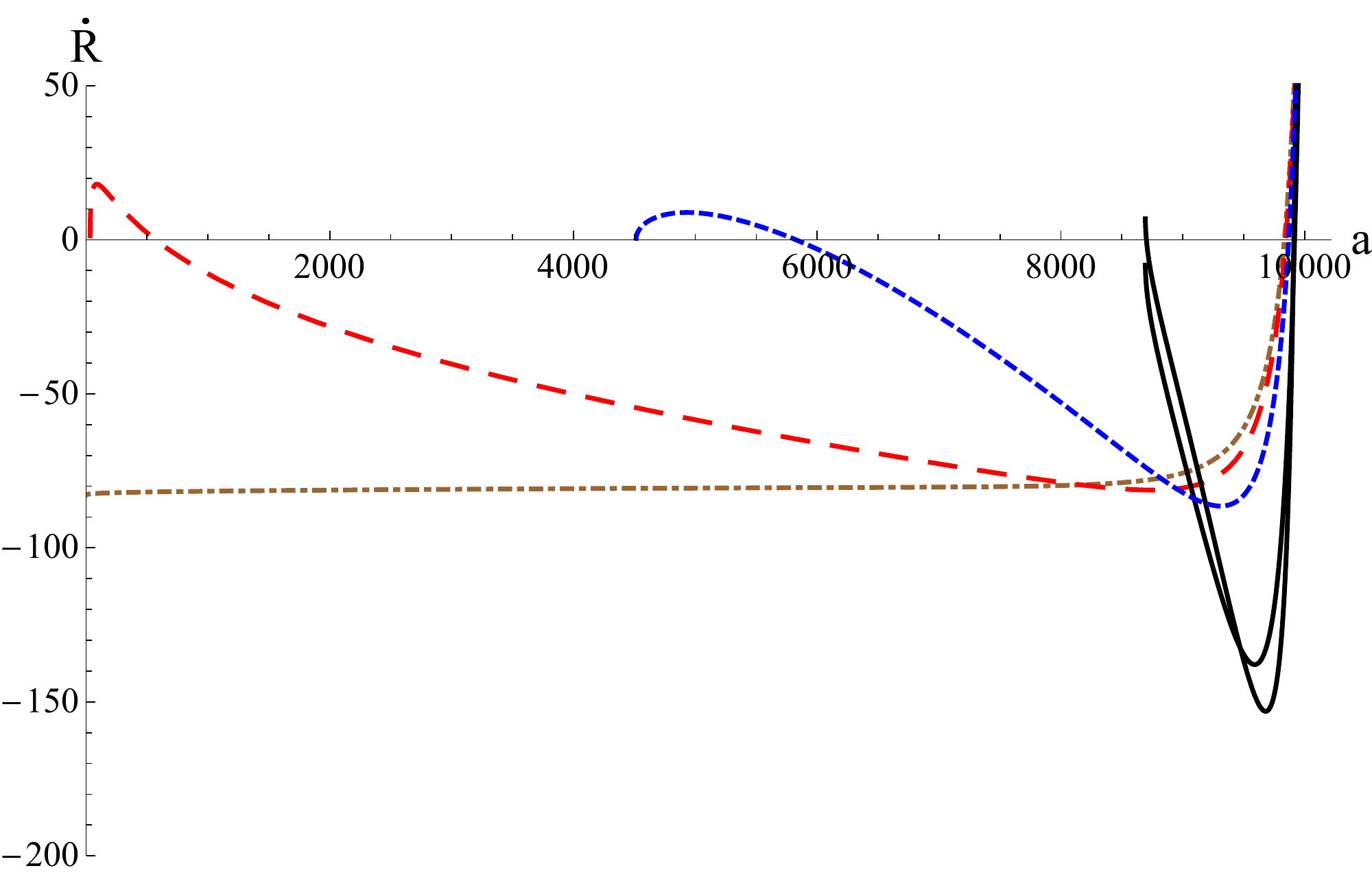} 
	\caption{In these graphs, we have plotted the curves for the Hubble rate and Ricci scalar as a function of the scale factor for the classical dynamics (dot-dashed), standard effective LQC dynamics (dashed), mLQC-I (solid) and mLQC-II (large-dashed). The values of the parameters chosen are $ A=-0.1, B=-1, a_o=10000 $ and $ \alpha=1/4.1 $. The weak singularity exists at $a=a_o$ for LQC and mLQC-II in both pre-bounce and post-bounce branches.}
	\label{fig:typeIV}
\end{figure}

If we choose $ 0 < \alpha < 1/2 $, then derivatives of curvature invariants diverge as $ a \rightarrow a_o $ while the energy density, pressure, Hubble rate and all the curvature invariants remain finite in the classical theory as well as all of the loop cosmological models under consideration.  But, the divergence in the time derivative of Ricci scalar is observed in both classical as well as loop cosmological models as shown in Fig. \ref{fig:typeIV}. The weak singularity occurs at $a = a_o$ both in the pre-bounce and post-bounce branches, as is the case for type-II singularity. However, unlike the case of type-II singularity, for type-IV singularity the evolution equation for $b$ does not break down as there is no divergence in pressure. Quantum geometry effects do not affect these divergences. 
Such a divergence which is only in the time derivative of curvature invariants is a weak singularity because the integral in eq.(\ref{krolak_condition}) is finite. Further, since the scale factor and Hubble rate at this singular event are finite, geodesics can be extended beyond type-IV singularity. 

For the above example, classically there is a big bang singularity in the past at which the Hubble rate and the Ricci scalar diverge. This is a strong singularity and is resolved in all the loop cosmological models and is replaced by a big bounce linking it to a past contracting universe. The type-IV singularity is present in these past (pre-bounce) branches as well in all three cases, except that the past branch is asymmetric in mLQC-I  while it is symmetric in LQC and mLQC-II. 

\section{Conclusions}

In this manuscript, we have analyzed the existence of various types of singularities in the effective dynamics of modified loop quantum cosmologies of spatially flat FLRW spacetime, namely mLQC-I and mLQC-II, and found the effective spacetimes to be geodesically complete and free from strong curvature singularities. For comparison, we have also included the analysis of classical and standard LQC dynamics. As discussed earlier, mLQC models result from quantizing the Lorentzian term in the Hamiltonian constraint independently of the Euclidean term \cite{Bojowald2002,Yang2009,Dapor2017}. In mLQC-II, the extrinsic curvature in the Lorentzian term is first replaced by the Ashtekar-Barbero connection before quantizing, while it is directly quantized in mLQC-I using Thiemann's identities involving holonomies of connection \cite{Thiemann}. Unlike LQC where the quantum constraint is a second order difference equation, these quantizations lead to fourth order difference equations \cite{von-Neumann}, resulting in a complicated effective phase space consisting of two branches of solutions each having its own set of Friedmann and Raychaudhuri equations. In mLQC-I, both branches are required to completely describe the phase space evolution generated by the effective Hamiltonian which leads to an asymmetric bounce \cite{Li2018}. On the other hand in mLQC-II, one of the branches turns out to be unphysical, leaving only one branch to describe the whole phase space evolution leading to a symmetric bounce similar to standard LQC \cite{Li2018b}.

Assuming the validity of effective spacetime description which has been verified extensively using numerical simulations \cite{numlsu}, we first showed that the energy density and Hubble rate are generically bounded in all of the considered loop cosmological models. The effective phase space is maximally extendible except for pressure divergences. Quantum geometric effects as understood in loop cosmologies still allow the Ricci scalar, the time derivative of the Hubble rate and the time derivative of Ricci scalar to diverge if the pressure or its time derivative diverges. Using analytical methods we showed that these divergences are harmless in all loop cosmological models as the geodesics can be extended beyond such divergent events and that these events do not lead to strong curvature singularities. We show that all strong curvature singularities are resolved in both mLQC-I and mLQC-II as is the case with standard LQC. 

Analytical results on geodesic completeness of effective spacetimes and absence of strong singularities are further understood via a phenomenological description of matter allowing various exotic singularities. In particular, choosing different values of parameters in the equation of state, type-I (big rip), type-II (sudden), type-III (big freeze), and type-IV (generalized sudden) singularities are studied. By numerical solving the effective dynamics in mLQC-I and mLQC-II we find that 
as in LQC \cite{ps09},  type-I and type-III singularities are resolved in both mLQC-I and mLQC-II. In contrast, the type-II and type-IV singularities are not resolved and are completely ignored by quantum geometry. Interestingly, type-I and type-III singularities are strong curvature singularities whereas type-II and type-IV are weak singularities and are harmless. Therefore, quantum gravitational effects as understood in loop cosmology for these spatially flat isotropic models seem to resolve only those singularities which are truly pathological. Whether the ignorance of weak singularities is a generic feature of loop cosmological models is not clear as the counter evidence of resolution of some weak singularities exists in spatially curved and anisotropic models \cite{psfv,kasner-flat}. Further, we should note that in quantum theory additional effects associated with quantum fluctuations can potentially soften or even resolve such weak singularities, as is the case for some models in Wheeler-DeWitt quantum cosmology \cite{type4q}. Finally, let us note that despite qualitative changes in the Planck scale physics of loop cosmological models, the main result of generic resolution of all strong singularities and geodesic completeness of effective spacetime earlier found for LQC \cite{ps09}, is found to be true in modified loop quantum cosmologies -- mLQC-I and mLQC-II.

\section*{Acknowledgments}
 This work is supported by  NSF grant PHY-1454832.

\end{document}